\documentclass[twocolumn,superscriptaddress,eqsecnum,pre,showpacs,showkeys]{revtex4}
\usepackage{amsmath,amssymb}
\usepackage{graphicx}

\begin{document}

\title{Towards the fundamentals of car following theory}
 \author{Ihor Lubashevsky}
 \affiliation{Theory Department, General Physics Institute, Russian Academy of
 Sciences, Vavilov Str. 38, Moscow, 119991 Russia}
 \affiliation{Institute of Transport Research, German Aerospace Center (DLR),
 Rutherfordstrasse 2, 12489 Berlin, Germany.}
 \author{Peter Wagner}
 \affiliation{Institute of Transport Research, German Aerospace Center (DLR),
 Rutherfordstrasse 2, 12489 Berlin, Germany.}
 \author{Reinhard Mahnke}
 \affiliation{Fachbereich Physik, Universit\"at Rostock, D--18051 Rostock, Germany}

\date{\today }

\begin{abstract}

  The problem of a car following a lead car driven with constant
  velocity is considered. To derive the governing equations for the
  following car dynamics a cost functional is constructed. This
  functional ranks the outcomes of different driving strategies, which
  applies to fairly general properties of the driver behavior.
  Assuming rational driver behavior, the existence of the Nash
  equilibrium is proved.
  Rational driving is defined by supposing that a driver corrects
  continuously the car motion to follow the optimal path minimizing
  the cost functional. The corresponding car-following dynamics is
  described quite generally by a boundary value problem based on the
  obtained extremal equations. Linearization of these equations around
  the stationary state results in a generalization of the widely used
  optimal velocity model.
  Under certain conditions (the ``dense traffic'' limit) the rational
  car dynamics comprises two stages, fast and slow. During the fast
  stage a driver eliminates the velocity difference between the cars,
  the subsequent slow stage optimizes the headway. In the ``dense
  traffic'' limit an effective Hamiltonian description is constructed.
  This allows a more detailed nonlinear analysis.
  Finally, the differences between rational and bounded rational
  driver behavior are discussed. The latter, in particular, justifies
  some basic assumptions used recently by the authors to construct a
  car-following model lying beyond the frameworks of rationality.

\end{abstract}
 \pacs{89.40.-a, 45.70.Vn, 02.50.Le}
 \keywords{car following model, motivated behavior, cost (utility) function,
           rationality, optimal driving conditions, Hamiltonian description of
           car dynamics}

\maketitle

\section{Car-following theories and basic properties of driver behavior}

Recently, the theoretical and empirical foundations of the physics of
traffic flow (for a review see Refs.~\cite{Chowdhury,Helbing}) has
come into the focus of the physical community. The motion of
individual cars has many peculiarities, since it is controlled by
motivated driver behavior, together with some physical boundaries.
Nevertheless, on macroscopic scales the vehicle ensembles displays
phenomena like phase formation and phase transitions widely met in
physical systems (see, e.g., Refs.~\cite{Chowdhury,Helbing,K99}). So,
the cooperative behavior of cars treated as active particles seems to
be of a more general nature than the mechanical laws and constructing
a consistent theory of traffic flow ``from scratch'' is up to now a
challenging problem.

To describe individual car dynamics a great variety of microscopic
models have been proposed. These models differ in the details of the
interaction between cars and the time update rule, ranging from
differential equations to cellular automata~\cite{Chowdhury,Helbing}.
There has been a big deal of work on the macroscopic behavior emerging
from the microscopic dynamics when exploring the behavior of systems
of interacting cars. However, there is still a lot of controversy in
both the macroscopic behavior when compared to reality
\cite{daganzo-critique-sync}, and in the microscopic foundations of
the individual car dynamics itself \cite{D1a}.

One currently adopted approach to specify the microscopic governing
equations of the individual car motion is the so-called social force
model. More details can be found in Refs.~\cite{SFM1,SFM2,SFM3}; here
only the basic ideas are touched on. At each moment $t$ of time, a
driver $\alpha$ changes the speed $v_{\alpha }$ of her car depending
on the road conditions and the arrangement of the neighboring cars:
\begin{equation}
\frac{dv_{\alpha }}{dt}=g_{\alpha }(v_{\alpha }) + \sum_{\alpha '\neq\alpha}
g_{\alpha \alpha^{\prime }}(x_{\alpha },v_{\alpha }\,|\,x_{\alpha ^{\prime
}},v_{\alpha'}). \label{int.1}
\end{equation}
The term $g_\alpha(v_\alpha)$ describes the motion of car $\alpha$ on
the empty road, whereas the term $g_{\alpha \alpha^{\prime
  }}(x_{\alpha },v_{\alpha }\,|\,x_{\alpha ^{\prime }},v_{\alpha
  ^{\prime }})$ allows for the interaction of car $\alpha $ with car
$\alpha ^{\prime }$ ($\alpha ^{\prime }\neq \alpha $). The interaction
is due to the necessity for driver $\alpha $ to keep a certain safe
headway distance between the cars. All the models mentioned above use
various \textit{Ans\"atze} for the last term.

The most interesting special case, which covers the majority of all
traffic flow situations, is that of single-lane traffic. Here, all
cars can be ordered according to their position on the road in the car
motion direction
$x_\alpha < x_{\alpha+1}$, here $\alpha=1,\ldots,N$.
%
%
Most models take into account solely nearest neighboring cars $\alpha $ and
$\alpha+1$, i.e., $g_{\alpha \alpha ^{\prime }}\neq 0$ for $\alpha ^{\prime
}=\alpha +1$ and, may be, $\alpha ^{\prime }=\alpha -1$ only. However, more
complicated models exist that can be described as models with anticipation
\cite{Wiedemann,TRANSIMS_TGF95,BL,K2,K2W} or the so-called intelligent driver
model \cite{IDM1,IDM2}.

The earliest ``follow-the-leader'' models \cite{FLM1,FLM2} relate the
acceleration $a_{\alpha}$ of car $\alpha$ to the velocity difference
$(v_{\alpha}-v_{\alpha+1})$ only, i.e.,
\begin{equation}\label{FLM}
    \frac{dv_{\alpha }}{dt} = -\frac{1}{\tau_v}(v_{\alpha}-v_{\alpha+1})\,,
\end{equation}
where $\tau_v$ is the characteristic time scale of the velocity
relaxation. In subsequent generalizations of this model $\tau_v$
became a function of the car motion state, in particular, of the
current velocity $v_\alpha$ and the headway
$h_\alpha=x_{\alpha+1}-x_\alpha-\ell$ (for a review see
Refs.~\cite{D1a,D1}).  Here, $\ell$ is the car length. In
Refs.~\cite{B1,B2} another approach called the optimal velocity model
is proposed that describes the individual car motion by
\begin{equation}
 \frac{dv_{\alpha }}{dt} = -\frac{1}{\tau_v} \left[ v_{\alpha } -
 \vartheta_{\text{opt}}(h_\alpha) \right]\,,
 \label{int.2}
\end{equation}
where $\vartheta_{\text{opt}}(h)$ is the steady-state velocity (the
optimal velocity) chosen by drivers as function of the headway $h$
between the cars. It should be noted that this approach is related to
much earlier safety distance models \cite{Ko1,Ko2,Newell}.

Concerning the fundamentals of approximations such as $a_\alpha =
a\left( v_\alpha, v_{\alpha+1},x_\alpha, x_{\alpha+1} \right)$, it is
noted that there are actually two stimuli affecting the driver
behavior. One of them is the necessity to move at the mean speed of
traffic flow, in the given case at the speed $v_{\alpha+1}$ of car
$\alpha+1$. So, first, driver $\alpha$ should control the velocity
difference $(v_{\alpha}-v_{\alpha+1})$. The other is the necessity to
maintain a safe headway distance $h_{\text{opt}}(v_\alpha)$ depending
on the current velocity $v_{\alpha}$. The ``following-the-leader''
models mainly take into account the former stimulus. The optimal
velocity model, conversely, allows for the latter stimulus only. More
sophisticated approximations, e.g.,
Refs.~\cite{IDM1,IDM2,Hel,Fritz,Xing,skPRE,H1} to name but a few,
allow for both stimuli. Note also a simple \textit{Ansatz} called the
combined model in Ref.~\cite{NWS} which is also related to the
intelligent driver model \cite{IDM1,IDM2}:
\begin{subequations}\label{IModels}
\begin{eqnarray}
    \nonumber
    \frac{dv_{\alpha }}{dt} &=&
    -\frac{(1-\kappa)}{\tau_v}\,[v_\alpha-v_{\alpha+1}]\\
    &&{}-\frac{\kappa}{\tau_v}\left[v_\alpha
    -\vartheta _{\text{opt}}(h_\alpha )\right] \,.
    \label{CM}
\end{eqnarray}
This equation takes into account both stimuli via a phenomenological
coefficient $0< \kappa < 1$. This is also the case for the Helly
model~\cite{Hel} which can be written as (cf.\ Ref.~\cite{D1a})
\begin{eqnarray}\nonumber
    \frac{dv_\alpha}{dt} & = &
    -\frac1{\tau_v}\,[v_\alpha-v_{\alpha+1}]\\
    &&{}+\frac1{\tau_v L_H}\left[
    (x_{\alpha +1}-x_\alpha)-h_{\text{opt}}(v_\alpha)\right] \,,
    \label{HelMod}
\end{eqnarray}
\end{subequations}
where $L_H$ is a certain spatial scale and $h_{\text{opt}}(v)$ is the optimal
headway distance chosen by drivers when moving at speed $v$. Ref.~\cite{Hel}
used a linear \textit{Ansatz} for $h_{\text{opt}}(v)$. Later~\cite{Xing}, this
model was generalized to allow for the dependence of the kinetic coefficients
on the motion state.

However, the question whether a collection of variables such as
$\left\{v_\alpha,v_{\alpha+1}, x_\alpha,x_{\alpha+1}\right\}$
specifies the acceleration $a_\alpha$ completely is not trivial.
Drivers are characterized by the motivated behavior rather than
physical regularities. For example, memory effects may be essential
%
%
and can destroy the direct relationship $a_\alpha =
a(v_\alpha,v_{\alpha+1}, x_\alpha,x_{\alpha+1})$.
%
%
Up to now, memory effects in car following modelling have been treated
only in a simplified version. This has been done by relating the
current acceleration $a_\alpha(t)$ to the velocities
$v_\alpha(t-\tau_a)$, $v_{\alpha+1}(t-\tau_a)$ and the headway
distance $h_\alpha(t-\tau_a)$ taken at a previous moment ($t-\tau_a$)
of time (for a review of such an approach concerning with the
``following-the-leader'' models see Refs.~\cite{D1a,D1}, regarding the
optimal velocity model see Refs.~\cite{D2,NNxxx,D2a}). Here $\tau_a$
is the formal delay time in the driver response which is treated as a
constant.
%
%
Such an approach, however, is rather formal, because, first, it is
not clear why the memory effects relate only two moments of time
instead of a certain interval as a whole. Second, a simple
physiological delay in the driver response as well as the mental
driver estimation of the surrounding situation should contribute to
the value of $\tau_a$. If the latter contribution is essential the
delay time $\tau_a$ is to depend substantially on the state of car
motion. Moreover,
%
%
from our point of view the memory effects stem from the fact that the
description of individual car motion is a boundary value problem
rather than an initial value problem. Indeed, at the current moment of
time the headway distance and the car velocity are quantities given
for the driver beforehand. Correcting the car motion she should choose
such a driving strategy that in a certain time interval the car
velocity and headway distance attain their optimal stationary values,
at least, approximately.

These memory effects are the topic of this paper. We propose that
drivers plan their behavior for a certain time in advance~\cite{we1}
instead of simply reacting to the surrounding situation. Similar ideas
related to the optimum design of a distance controlling driver
assistance system are discussed in Ref.~\cite{q}. Mathematically, the
driver's planning of the further motion is just to find extremals of a
certain priority functional that ranks outcomes of different driving
strategies.

The derivation of microscopic governing equations for systems with
motivated behavior based on a certain ``optimal self-organization''
principle has been discussed recently
\cite{Helbing,Ac3,MSO1,MSO2,MSO3,MSO4}. The assumption adopted in
these works is that individuals try to minimize the interaction
strength or, equivalently, to optimize their own success and to
minimize the efforts required for this. The approach discussed here
applies to the concepts of mathematical economics, namely, to the
notion of preferences and utility (see, e.g., Ref.~\cite{Ut}).  In
Ref.~\cite{we1} a specific form of the priority functional has been
proposed. From that, a certain \textit{Ansatz} like the combined
model~(\ref{CM}) or the Helly model~(\ref{HelMod}) could be
\textit{derived}. Nevertheless, the question how to find the priority
functional ``from scratch'' remains open.

In the present paper the priority functional is constructed by
applying to general properties of the driver behavior for a simplified
situation. A car following (no overtaking allowed) a lead car which is
driven with constant speed $V$ is considered. The task is to derive
governing equations for the following car motion specified by the time
dependence of the velocity $v(t)$ and the headway distance $h(t)$.

\section{The cost function of driving\label{sec:1}}

\subsection{General properties of driver preference}

Assuming a driver to be able of comparing any two states
$\{h_{1},v_{1}\}$, $\{h_{2},v_{2}\}$, the phase plane $\{h>0,v>0\}$
can be ordered by a preference relation $\preceq$. Therefore, the
existence of a cost function $\mathcal{F}(h,v)$ may be assumed such
that:
%
%
\begin{equation}
 \{h_{1},v_{1}\}\preceq\{h_{2},v_{2}\}\Leftrightarrow\mathcal{F}
 (h_{1} ,v_{1})\geq\mathcal{F}(h_{2},v_{2})\,.
 \label{1.1}
\end{equation}
Typically, $-\mathcal{F}(h,v)$ is called the utility function and (for
it) the relation $\preceq$ matches the inequality $\leq$. Here, the
use of condition~(\ref{1.1}) is preferred, because then the cost
function $\mathcal{F}(h,v)$ is similar to the free energy of physical
systems, with the minima as the stationary states.
%
%
%

Obviously, the cost function $\mathcal{F}(h,v)$ cannot be specified
completely because a composite function
$\Psi\left[\mathcal{F}(h,v)\right]$ also meets condition~(\ref{1.1})
for any increasing function $\Psi\left[\cdot\right]$.
%
%
Therefore, at the current stage of the theory development any
approximation or \textit{Ansatz} adopted for the cost function
$\mathcal{F}(h,v)$ has no meaning.

To overcome this problem and to fix a certain class of the cost functions the
preference relation defined on a set of car motion paths $\{h(t),v(t)\}$ is
considered. A car moving from one point to another can meet different states of
traffic flow. It enables one to unite various points on the phase plane in an
element treated as a whole with respect to the driver preference, reducing the
numerical uncertainty in their evaluation. These parts are supposed, first, to
connect the same origin-destination pair $\{O,D\}$ and, second, to go through a
traffic flow pattern whose structure is known to the driver beforehand. Such a
set obviously can be ordered by the driver preference and the corresponding
cost functional $\mathcal{L}\{h(t),v(t)\}$ is assumed to exist. Here, only the
fact that a car motion path can consist of various fragments of the phase plane
is concerned with. Analysis of the real car dynamics is postponed to the
following section.

\begin{figure}
\begin{center}
\includegraphics[width=7cm]{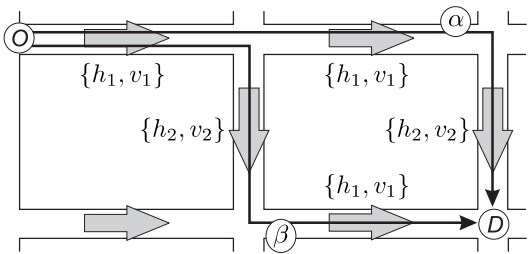}
\end{center}
\caption{Two equivalent car motion paths $\alpha$ and $\beta$ illustrating the
  property of the cost functional with respect to path partitioning.
  Letters $O$ and $D$ stand for the origin and destination points,
  respectively, and the symbols $\{h_i,v_i\}$ denote the local states
  of traffic flow.\label{F2}}
\end{figure}

Figure~\ref{F2} shows two paths $\alpha$ and $\beta$ which are equivalent with
respect to the driver preference. They differ from each other only in the order
of passing through the traffic states, $\{h_{1},v_{1}\}\to \{h_{1},v_{1}\}\to
\{h_{2},v_{2}\}$ for the path $\alpha$ and $\{h_{1},v_{1}\}\to
\{h_{2},v_{2}\}\to \{h_{1},v_{1}\}$ for the path $\beta$.  It demonstrates the
fact that if a path goes through a certain small neighborhood of the point
$\{h,v\}$ one or many times then only the cumulative time interval $\delta t$
during which the system was located in the given neighborhood contributes to
the cost functional. The corresponding weight coefficient $\mathcal{F}(h,v)$
obviously plays the role of a certain cost function for the given traffic
state.  The resulting cost functional for the whole path is
\begin{equation}
   \mathcal{L}\left\{h(t),v(t)\right\}  = \int\limits_{{\text{origin}}}^
   {{\text{destination}}}\mathcal{F} \left[h(t),v(t)\right]dt\,.
   \label{1.2}
\end{equation}
If the driver has incomplete information about the traffic flow pattern ahead
she plans the further motion only within some limits, spatial and temporal
ones. This can be taken into account by introducing a certain cofactor in front
of the function $\mathcal{F}(h,v)$ which depends on spatial coordinates or
time.

%
%
The initial introduction of the cost function remains the freedom of its choice
too ``wide'', as was noted above. The constructed form of the cost functional
reduces this freedom substantially. Namely, let ${\mathcal{F}}(h,v)$ and
$\widetilde{\mathcal{F}}(h,v)$ be two functions entering a cost functional
similar to (\ref{1.2}) with the prediction effect such that they preserve the
preference relation with respect to the car motion paths on an one-lane road.
Then, as shown in Appendix~\ref{Appp:A}, these cost functions are related by
the expressions
\begin{subequations}
\begin{eqnarray}
    \label{cf.4}
    \widetilde{\mathcal{F}}(h,v) & = & \varkappa \mathcal{F}(h,v) + C_a(h)\,v\,,\\
    \label{cf.11}
    \widetilde{\mathcal{F}}(h,v) & = & \varkappa \mathcal{F}(h,v) + C_b(t)\,,
\end{eqnarray}
\end{subequations}
depending on the spatial or temporal limitation in the driver prediction. Here
$\varkappa >0$ is an arbitrary positive constant and $C_a(h)$, $C_b(t)$ are
some functions of the headway distance $h$ or the time $t$. In other words,
relations~(\ref{cf.4}), (\ref{cf.11}) specify the family of the admissible type
of cost functions preserving the driver preference relation and, in addition,
correspond to linear transformation of the cost functional
(Appendix~\ref{Appp:A}).

For example, $\left[\mathcal{L}\{ h(t)\}\right]^{2}$ is also a cost functional.
However, formula~(\ref{1.2}) enables us to fix its specific form being linear
with respect to time integration along car motion paths. We note that this
construction conforms with regarding the motion cost as the ``physiological''
cost of driving per time, i.e., having the meaning of a rate.

Both of these families possess the same freedom in choosing a specific form of
the cost function especially dealing with the extremals of the cost functional.
In fact, the type~(\ref{cf.11}) is chosen in the following.
%

\subsection{Characteristic features of the car motion cost}

In the following, the priority function of the car motion state with
respect to the velocity $v$ and the reciprocal value $\rho = 1/h$ of
the headway distance will be characterized.
%
%
Note, that the value $\rho$ is not the real car density on a highway. The
present analysis ignores the car length, so we have preferred to use the value
$\rho$ defined as above.
%
%
The most preferable state is the motion on an empty road ($\rho = 0$) at
maximum speed $\vartheta_{\text{max}}$. This maximum speed is determined by
external factors. So, at $\{\rho=0,v=\vartheta_{\text{max}}\}$ the cost
function $\mathcal{F}\left(\rho,v\right):=\left.  \mathcal{F}\left( h,v\right)
\right\vert_{h=1/\rho}$ has its global minimum:
\begin{equation}
    \mathcal{F}(0,\vartheta_{\text{max}})=0\,,
    \label{1.3}
\end{equation}
set equal to zero keeping in mind the aforesaid about the freedom in
specifying the cost function. Since there is no other minimum for the
motion on an empty road,
\begin{equation}
    \mathcal{F}(0,0)=1\,
    \label{1.4}
\end{equation}
can be fixed.

Because of the car construction a driver can visually control the
headway distance within some value $l\sim 1$--2~m. When the headway
distance $h$ attains such values a driver has to stop her car because
of possible collision even at sufficiently slow velocities. So, $l$
coincides with the characteristic headway distance in dense jams where
the car density attains the possible maximum. In the following, all
headway distances are related to the scale $l$ and the dimensionless
variable $\rho l$ is used.

When not moving at all ($v=0$) and the density of cars surrounding the
given car does not come close to the limit values, i.e., $\rho l\ll
1$, it does not matter how many cars are located in the vicinity. So
the assumption that at $v=0$ and for $h\gg l$ the cost function is
independent of $\rho$ can be used:
\begin{equation}
    \mathcal{F}(\rho,0)=1\quad \text{for}\quad \rho l\ll1\,.
    \label{1.5}
\end{equation}
In all other cases, $\mathcal{F}\left(\rho,v\right)$ depends on
certain combinations of $\rho$ and $v$ rather than on $\rho$
individually, at least when $\rho l \ll 1$.

Considering the behavior of $\mathcal{F}\left(\rho,v\right)$ for a
fixed speed $v$ it can be stated that driving with small values of $h$
(large values of $\rho$) requires a lot of effort. Thus,
$\mathcal{F}\left(\rho,v\right)$ increases with $\rho$. The opposite
case of small values of $\rho$ (large values of $h$) deserves special
attention. Without the possibility of overtaking no especially
attractive headway distance can be marked.
Therefore, we assume that the cost function $\mathcal{F}\left(
  \rho,v\right)$ possesses the only one minimum attained at the
boundary point $\rho=0$ provided the velocity $v$ is fixed.

Keeping in mind the aforesaid it is reasonable to write
\begin{equation}\label{1.6na}
 \mathcal{F}\left(  \rho,v\right)  =\mathcal{F}\left(  0,v\right) +
 f\Big(\frac{v}{\vartheta_{\text{max}}}\Big)(\rho l)^m
\end{equation}
for $\rho l\ll 1$. Here the exponent $m$ is a constant and the
function $f(z)\to 0$ as $z\to 0$. Since the cost function attains its
minimum at the
%
%
boundary point we may set $m=1$.
Moreover, the latter term on the right-hand side of Exp.~(\ref{1.6na})
can be interpreted as a certain ``interaction'' potential between the
following and lead cars which is long-distant one for $m=1$. A
detailed analysis of effects caused by the value of the exponent $m$
requires an individual study. Here the value $m = 1$ is actually
chosen as just a simple reasonable assumption.
%
%
Since the effect of the surrounding cars is depressed for small values
of the car velocity $v$ the function $f(z)$ is also to attain its
minimum at $z=0$.  Therefore, $f(z)=z^2$ as $z\ll1$ should hold. In
other words, inside a certain neighborhood of the origin $\{\rho = 0,
v=0\}$ the following expansion
\begin{equation}
 \mathcal{F}\left(\rho,v\right)  = \mathcal{F}\left(0,v\right)
 +\frac{v^2}{\vartheta_{\text{max}}^2}\,\rho l
 \label{1.6n}
\end{equation}
can be adopted.

\begin{figure}
\begin{center}
\includegraphics{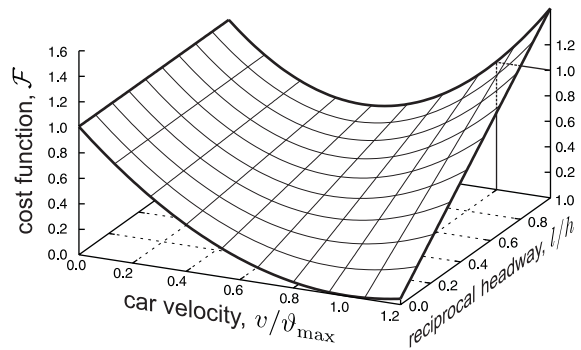}
\end{center}
\caption{Characteristic form of the cost function, Eq.~(\ref{1.7}).\label{F3}}
\end{figure}

%
%
Taking into account these speculations about the behavior of the cost
function $\mathcal{F}(\rho,v)$ caused by variations of both its
arguments, the following simple \textit{Ansatz} will be used
subsequently:
%
%
\begin{equation}
  \mathcal{F}\left(  \rho,v\right)  =\left(  1-\frac{v}{\vartheta_{\text{max}}
  }\right)  ^{2}+\frac{v^{2}}{\vartheta_{\text{max}}^{2}}\,\rho l\,.
  \label{1.7}
\end{equation}
Figure~\ref{F3} displays this function. It has only one global minimum
at $\rho = 0$ and $v=\vartheta_{\text{max}}$. For a fixed velocity $v$
it attains a local minimum at the boundary $\rho = 0$.
\textit{Ansatz}~(\ref{1.7}) generalizes the adopted assumptions about
the cost function. The former term takes into account
relations~(\ref{1.3}), (\ref{1.4}), the latter one is based on
approximation~(\ref{1.6n}). Of course, the parabolic
\textit{Ansatz}~(\ref{1.7}) is approximate only in the limit $\rho
l\ll 1$. Since this is the main region of interest, this does not
constrain its usefulness.

To deal with the car dynamics we should construct the cost functional
$\mathcal{L}\left\{h(t)\right\}$ for the car motion paths $\left\{
h(t)\right\}$ treated now as continuous functions of time $t$. We note that the
time dependence of headway distance $h(t)$ gives us the complete information
about the car dynamics due to the relationship $dh/dt=V-v$. Leaping ahead, we
say that facing this problem it is necessary to introduce additional notions.
First, we should expand the phase space in describing the car motion state
because transient processes are now the subject of consideration.  Second,
drivers plan their behavior for a certain time in advance instead of simply
reacting to the surrounding situation. So we should specify the region inside
which a driver can monitor the traffic flow evolution and, thus, plan driving
her car. A priority functional similar to Eq.~(\ref{1.2}) must span the time
interval corresponding to this region. Beyond it the contribution of the path
fragments to evaluating the car motion quality at the given moment of time has
to be fairly minor.

\section{Rational dynamics of car motion\label{sec:2}}

\subsection{Cost functional and the extremal equation\label{sec:ee}}

Dealing with transient processes in the car motion we should consider
once more the collection of phase variables characterizing the cost of
car motion at the current moment of time. Keeping in mind conventional
driver experience, we will expand the current state of car motion
given by headway $h$ and velocity $v$ with the car acceleration $a$.
This is essential because a
driver cannot change the position and velocity of her car immediately,
they vary continuously in time and contain no sharp jumps. Conversely,
a driver controls the acceleration directly governing the car motion.
Besides, she can change the acceleration practically without delay
because in the present analysis it is quite reasonable to ignore time
scales related to physiological properties of the driver or to the
mechanical properties of the car. So, the cost function
$\mathcal{F}^{d}\left(h,v,a\right)$ for the motion state
$\left\{h,v,a\right\}$ can be written:
\begin{equation}
    \mathcal{F}^{d}\left(  h,v,a\right)  =\mathcal{F}\left(  h,v\right)
    +\frac{\tau^{2}a^{2}}{\vartheta_{\text{max}}^{2}}\,,
    \label{2.1}
\end{equation}
where the time scale $\tau\gtrsim 1$~s characterizes the acceleration
capability of the car. In writing this expression we have assumed that driving
without acceleration is preferable. Then, the cost function
$\mathcal{F}_{d}\left(h,v,a\right)$ had been expanded into a Taylor series with
respect to $a$, keeping the leading term only. According to the result to be
obtained the time scale $\tau$ entering expression~(\ref{2.1}) and the one used
in equation~(\ref{CM}) are practically the same. It should be noted that
confining ourselves to expansion~(\ref{2.1}) with respect to $a$ the difference
between acceleration and deceleration processes has been lost. In reality, they
are different. Ignoring this difference leads to models where cars can crash.
It is possible to take into account this effect using the approach under
development, which, however, is worthy of individual investigation and will be
done somewhere else.

\begin{figure}
  \includegraphics{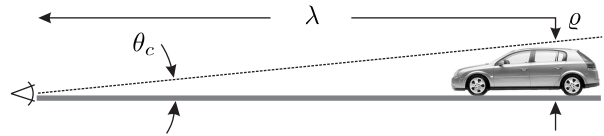}
  \caption{Illustration of the recognition distance $\lambda$ and its
    relationship with the limit angle of perception
    $\theta_c$.\label{F4}}
\end{figure}

For a real driver, various thresholds in the driver recognition of
hazards and obstacles exist \cite{Rec1}. One of them is the distance
$\lambda$ at which a driver can recognize the behavior of other
objects. This distance is usually related to the threshold of the
visual angle $\theta_c$ subtended, for example, by vehicles ahead
(Fig.~\ref{F4}). The value of $\theta_c$ can be estimated as
$\theta_c\sim 15$--30~min of arc \cite{Rec2}. The critical angle
$\theta_c$, the characteristics height $\varrho$ of cars, and the
corresponding mean distance $\lambda$ are related by
\begin{equation*}
    \theta_c \sim \frac{\varrho}{\lambda}\,.
\end{equation*}
This allows for the estimation of the recognition distance as
$\lambda\sim 200$--400~m setting $\varrho\sim 2$~m and $\theta_c\sim
30$--15~minutes of arc. As previously~\cite{we1}, we relate the driver
anticipation with the region of size $\lambda$ in front that is
clearly observable and where she can recognize the car behavior. A
driver plans her motion based on the information received by
monitoring traffic flow inside the observable region. Under normal
conditions this region should enable her to govern the motion
effectively, for example, to decelerate in advance avoiding a possible
accident. Therefore its size $\lambda$ has to meet the inequality
$\lambda\gtrsim\vartheta_{\text{max}}\tau$. Leaping ahead, we
introduce the value
\begin{equation}
 \sigma = \frac{\vartheta_{\text{max}}\tau}{\lambda} \ll 1
 \label{2.sigma}
\end{equation}
treated as a small parameter in the theory to be developed. In
particular, for $\tau \approx 1$~s, $\vartheta_{\text{max}} \approx
100$~km/h, and $\lambda \approx 300$~m we have $\sigma\approx 0.1$

%
%
The size $\lambda$ of the recognition distance can be estimated using
another argumentation. If $\tau_d\sim 10$~s is the typical
deceleration time from the velocity $\vartheta_{\text{max}}$ attained
on the given empty road to zero value then the estimate $\lambda\sim
\vartheta_{\text{max}}\tau_d$ should hold.  For
$\vartheta_{\text{max}}\sim 100$~km/h we have again $\lambda \approx
300$~m.
%
%

Now the cost functional of car motion can be written in an integral
form as expression~(\ref{1.2}) containing the integrand~(\ref{2.1}).
The car is assumed to be located at point $x$ along the road and to
move with speed $v$ at the current moment $t$ of time. The possible
paths of further motion $\{\mathfrak{r}(\mathfrak{t},t),\
\mathfrak{t}\geq t\}$ form the set $\mathfrak{S}(t,x,v)$ on which the
cost functional is defined. Here and below Gothic letters will be used
to label the path variables. The function
$\{\mathfrak{r}(\mathfrak{t},t)\}$ allows for a derivation of all the
dynamical variables, the headway distance
$\mathfrak{h}(\mathfrak{t})$, the car velocity
$\mathfrak{u}(\mathfrak{t})$, and the acceleration
$\mathfrak{a}(\mathfrak{t})$ of the trial car path.
%
%
The cost functional evaluating the quality of the path
$\{\mathfrak{r}(\mathfrak{t},t)\}$ can be written as
\begin{subequations}
\begin{align}\label{2.4n}
    \mathcal{L}\{\mathfrak{r}\} & =
    \int_t^\infty \exp\left[-\frac{\mathfrak{r}(\mathfrak{t})-x}{\lambda}
    \right]
    \mathcal{F}^d(\mathfrak{h},\mathfrak{u},\mathfrak{a})\,d\mathfrak{t}\,,\\
\intertext{or}
    \mathcal{L}\{\mathfrak{h}\} & =
    \int_t^\infty \exp\left[-\frac{V}{\lambda}(\mathfrak{t}-t) \right]
    \mathcal{F}^d(\mathfrak{h},\mathfrak{u},\mathfrak{a})\,d\mathfrak{t}\,.
    \label{2.7}
\end{align}
\end{subequations}
Forms~(\ref{2.4n}) and (\ref{2.7}) correspond to the driver prediction
with spatial or temporal limitations, respectively. The cost
functional of form~(\ref{2.4n}) was used in \cite{we1}. Subsequently,
functional~(\ref{2.7}) is used because its form simplifies the
mathematical manipulations when the velocity $V$ of the lead car is
fixed. In particular, dealing with functional~(\ref{2.7}) we can
specify directly the set of trial paths $\mathfrak{S}(t,h,v)$ using
solely the time dependence of $\{\mathfrak{h}(\mathfrak{t},t)\}$ of
the headway distance. The path variables
$\mathfrak{u}(\mathfrak{t},t)$, $\mathfrak{a}(\mathfrak{t},t)$ are
completely determined by the dependence
$\{\mathfrak{h}(\mathfrak{t},t)\}$:
\begin{equation}
 \mathfrak{u}(\mathfrak{t},t)=V-\frac{\partial\mathfrak{h}
 (\mathfrak{t},t)}{\partial\mathfrak{t}}\,,\quad\mathfrak{a}
 (\mathfrak{t},t)=-\frac{\partial^{2}\mathfrak{h}(\mathfrak{t},t)}
 {\partial\mathfrak{t}^{2}}\,.
 \label{2.4}
\end{equation}
Furthermore, the cost functional~(\ref{2.7}) matches the driver
prediction with temporal limitation, the mechanism which also allows
for the interpretation of the recognition region size in terms of
$\lambda\sim \vartheta_{\text{max}}\tau_d$. However, both of the cost
functionals lead practically to the same results for the analyzed
situation.
%
%

Each one of trial paths originates at time $t$ and starts from $\left(
  h,v\right)$ on the phase plane determined by the headway distance
and velocity of the car at the current time $t$. Since we investigate
the car motion inside traffic flow but not processes of leaving it we
assume that time variations $h(t)$ of the headway distance are
bounded. Therefore the trial paths fulfil
\begin{equation}
 \left.\mathfrak{h}(\mathfrak{t},t)\right\vert _{\mathfrak{t}=t}=h(t)\,,
 \quad
 \left. \mathfrak{u}(\mathfrak{t},t)\right\vert_{\mathfrak{t}=t}=v(t)
 \label{2.5}
\end{equation}
and do exhibit bounded variations only as time goes to infinity. In other
words, there is a constant $C>0$ with
\begin{equation}
  \left\vert \mathfrak{h}(\mathfrak{t}, t)\right\vert <C\quad\text{for}\quad
  \mathfrak{t}> t \,.
  \label{2.6}
\end{equation}

In what follows, the driver behavior will be described by the
extremals of the cost functional~(\ref{2.7}).  Using the standard
variational technique and taking into account conditions~(\ref{2.5}),
(\ref{2.6}) the governing equation for these extremals can be derived:
\begin{multline}
    \frac{d^{2}}{d\mathfrak{t}^{2}}\partial_{\mathfrak{a}}\mathcal{F}^{d}
    -2\frac{V}{\lambda}\frac{d}{d\mathfrak{t}}\partial_{\mathfrak{a}}
    \mathcal{F}^{d}+\frac{V^{2}}{\lambda^{2}}\partial_{\mathfrak{a}}
    \mathcal{F}^{d}\\
    -\frac{d}{d\mathfrak{t}}\partial_{\mathfrak{u}}\mathcal{F}
    ^{d}+\frac{V}{\lambda}\partial_{\mathfrak{u}}\mathcal{F}^{d}-\partial
    _{\mathfrak{h}}\mathcal{F}^{d}=0\,.
    \label{2.8}
\end{multline}

To study the spectral properties of Eq.~(\ref{2.8}) it is linearized
around the stationary solution $(h_{V},V)$. Its eigenfunctions can be
found by the \textit{Ansatz}
\begin{equation*}
    \mathfrak{h}_{\zeta}(\mathfrak{t}) \propto
    \exp\left(-\zeta \frac{\mathfrak{t}}{\tau}\right)  \,.
\end{equation*}
The obtained eigenvalue equation is given by
\begin{equation}
    \left(  \zeta+\phi\right)  ^{2}\zeta^{2}-\Lambda\left(  \zeta+\phi\right)
    \zeta+\frac{1}{4}\Omega=0\,,
    \label{2.800}
\end{equation}
with the coefficients
\begin{gather}
    \phi  =\frac{V}{\vartheta_{\text{max}}}\,\sigma\,,\qquad
    \Lambda   =\frac{1}{2}\vartheta_{\text{max}}^{2}\partial_{v}^{2}
    \mathcal{F}>0\,,
    \label{2.802}\\
    \Omega   =2\tau^{2}\vartheta_{\text{max}}^{2} \left(\partial_{h}
    ^{2}\mathcal{F}-\frac{V}{\lambda}\partial_{h}\partial_{v}\mathcal{F}
    \right)
    \label{2.803}
\end{gather}
and the derivatives are taken at the stationary point $(h_{V},V)$. The
inequality $\partial_{v}^{2}\mathcal{F}>0$ is supposed to hold
beforehand, in particular, it is the case for the cost
function~(\ref{1.7}). Equation~(\ref{2.800}) possesses four roots, one
pair $\left\{\zeta_{+},\zeta_{-}\right\}$ of them have positive real
parts, the other
$\left\{\zeta_{+}^{\prime},\zeta_{-}^{\prime}\right\}$ have negative
ones. Since the eigenfunctions with the eigenvalues
$\left\{\zeta_{+}^{\prime},\zeta_{-}^{\prime}\right\}$ diverge as
$\mathfrak{t} \rightarrow\infty$ we must omit them by virtue of
condition~(\ref{2.6}).
%
%
Roughly speaking, these divergent eigenfunctions describe the process
of a driver leaving traffic flow, for example, to stop the car. Such
processes are not under consideration.
%
%
The former pair of eigenvalues are given by:
\begin{equation}
    \zeta_{\pm}=-\frac{1}{2}\phi+\left[  \frac{1}{4}\phi^{2}+ \frac{1}{2}\Lambda
    \pm\frac{1}{2}\sqrt{\Lambda^{2}-\Omega}\right]^{\tfrac{1}{2}}.
    \label{2.804}
\end{equation}
So, in a small neighborhood of the stationary point $(h_{V},V)$ of
equation~(\ref{2.8}) any extremal
$\mathfrak{h}_{\text{opt}}(\mathfrak{t},t)$ can be written as
\begin{eqnarray}\nonumber
    \mathfrak{h}_{\text{opt}}(\mathfrak{t},t)= h_{V} &+& h_{+}\exp\left(-\zeta
    _{+}\frac{\mathfrak{t}-t}{\tau}\right) \\
    {}&+ &h_{-}\exp\left(  -\zeta_{-}\frac{\mathfrak{t}-t}{\tau}\right)  \,,
    \label{2.56}
\end{eqnarray}
where $h_{+}$ and $h_{-}$ are some constants.

Equation~(\ref{2.8}) is of fourth order, so its general solution is
specified by four conditions. Solutions that diverge as time goes to
infinity have to be omitted (they are related to
$\left\{\zeta_{+}^{\prime},\zeta_{-}^{\prime}\right\}$). Their
divergence is due to the exponential cofactor in integral~(\ref{2.7})
so it is retained beyond a small neighborhood of the stationary point
$(h_V,V)$. Therefore, only two conditions are needed to specify the
desired solution of Eq.~(\ref{2.8}). In particular, the initial
headway distance $h$ and the car velocity $v$ determine it completely.
The latter statement can also be proved for the more general
form~(\ref{2.4n}) of the cost functional \cite{we1}.

\subsection{Rational driver behavior and Nash equilibrium}

\begin{figure}
  \includegraphics{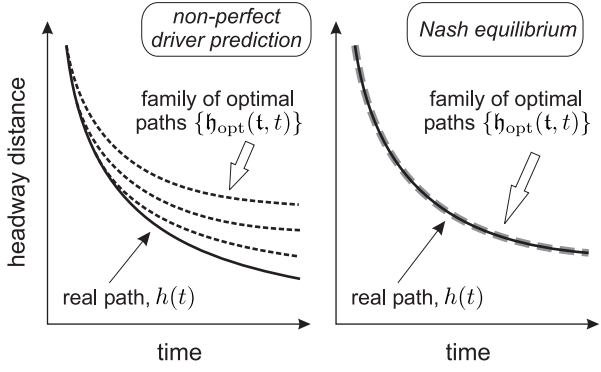}
  \caption{Illustration of rational driver strategy. The left panel
    corresponds to the case where the driver's estimate of the car
    motion quality is not perfect. This is described by the cost
    function $\mathcal{F}(\mathfrak{h},\mathfrak{u},\mathfrak{t}\mid
    t)$ containing the current time $t$ as its argument. So the found
    optimal paths $\mathfrak{h}_{\text{opt}}(\mathfrak{t},t)$ depend
    on the time $t$ when the driver plans the further motion.
    The right panel presents the ideal case where the driver is able
    to evaluate the motion cost precisely and the corresponding cost
    function $\mathcal{F}(\mathfrak{h},\mathfrak{u},\mathfrak{t})$
    does not contain the time $t$. So, the optimal path family
    degenerates into one curve. It should be noted that the latter is
    correct even if the time $\mathfrak{t}$ enters the cost function
    via variations of the lead car velocity.
    \label{FN}}
\end{figure}

In the previous section the dynamical cost functional was stated. At
the next step the driver strategy based on this evaluation of the
motion quality should be described (Fig.~\ref{FN}). Following
\cite{we1}, it is supposed that the driver, first, in planning the
further motion chooses the path
$\mathfrak{h}_{\text{opt}}(\mathfrak{t}, t)$ minimizing the motion
cost functional~(\ref{2.7}),
\begin{equation}
 \text{driver:}\rightarrowtail\mathfrak{h}_{\text{opt}}(\mathfrak{t},t)
 \Rightarrow\min_{\mathfrak{h}(\mathfrak{t}, t)\in\mathfrak{S}(t,h,v)}
 \mathcal{L}\left\{\mathfrak{h}(\mathfrak{t}, t)\right\} \,.
 \label{2.60}
\end{equation}
To do this the driver ``solves'' equation~(\ref{2.8}) subject to
conditions~(\ref{2.5}), (\ref{2.6}) and, thus, get the optimal path of
further driving, $\mathfrak{h}_{\text{opt}} (\mathfrak{t},t)$. As
noted in the previous section the optimal path choice is completely
determined by the current car velocity $v$ and the headway distance
$h$. The terminal conditions, i.e., the goal of reaching the
steady-state motion is implied. Since the driver controls the car
motion through choosing the adequate value $a$ of acceleration she has
to ``find'' the second derivative of
$\mathfrak{h}_{\text{opt}}(\mathfrak{t},t)$ with respect to the former
argument $\mathfrak{t}$ (see Exp.~(\ref{2.4})) and, then, to
``calculate'' the result at the current time. Therefore,
\begin{equation}
a(t)=-\lim_{\mathfrak{t}\rightarrow t+0}
\frac{\partial^{2}\mathfrak{h}_{\text{opt}}(\mathfrak{t},t)}{\partial\mathfrak{t}^{2}}\,,
    \label{2.61}
\end{equation}
relates the current car acceleration $a$ to the current headway
distance $h$ and the car velocity $v$.

Formula~(\ref{2.61}) describes the driver's choice at the current
moment $t$ of time. To convert it into the governing equation of car
motion we adopt the second assumption that the driver performs this
choice continuously:
\begin{equation}
a = \mathcal{R}(h,v,V)\,.    \label{2.62}
\end{equation}
All the car following models based on this equation, but with
different particular forms of the cost function $\mathcal{F}^d\left(
  \mathfrak{h},\mathfrak{u},\mathfrak{a}\right)$ may be categorized as
the rational car dynamics approximation.

Equation~(\ref{2.62}) holds even if the cost function
$\mathcal{F}\left(\mathfrak{h},\mathfrak{u},\mathfrak{t}\mid t\right)$
depends explicitly on the time $t$ at which the driver plans her
further motion.  Then, the driver evaluation of traffic flow state
will change in time. Therefore, the resulting path $\{h(t)\}$ of the
real car motion envelops the family of the optimal paths
$\{\mathfrak{h}_{\text{opt}}(\mathfrak{t},t)\}$ generated at different
moments $t$ of time (Fig.~\ref{FN}, left panel).

The present paper assumes driving to be perfect. The driver's choice
is based on the precise knowledge of the cost function
$\mathcal{F}^d\left( h,v,a,t\right)$ depending solely on the current
headway distance $h$, car velocity $v$, acceleration $a$, and, may be,
the current time $t$. It does not depend on the moment of time when
the driver evaluates the possible paths of her further motion. This
assumption leads to what is called the Nash equilibrium in game theory
\cite{NE}. If, at time $t_{0}$ the driver has found the optimal path
$\mathfrak{h}_{\text{opt}}(\mathfrak{t},t_{0})$ of the car motion
(which meets equation~(\ref{2.8}) and does not contain explicitly
$t_{0}$), then any re-computation a certain time $\mathfrak{t}>t_{0}$
later gives the same result, $\mathfrak{h}_{\text{opt}
}(\mathfrak{t},t) = \mathfrak{h}_{\text{opt}}(\mathfrak{t}, t_{0})$
for $\mathfrak{t}\geq t$. In other words, if at time $t_{0}$ the
driver has chosen an optimal path
$\mathfrak{h}_{\text{opt}}(\mathfrak{t}, t_{0})$ then the further
motion will be described by it independently of either the driver
follows it without correction or optimizes the car motion continuously
(Fig.~\ref{FN}, right panel).

The Nash equilibrium in the driver strategy enables us to conclude that
equation~(\ref{2.8}) describes the real car dynamics, not only the imaginary
paths existing in the driver's mind during her planing of the further motion.
In particular, it can be rewritten in a form containing the real
acceleration~$a$, velocity $v$, and headway distance $h$:
\begin{multline}
    \left(
    \frac{d^{2}a}{dt^{2}}-2\frac{V}{\lambda}\frac{da}{dt}+\frac{V^{2}}{\lambda
    ^{2}}a \right)
    \\
    -\frac{\vartheta_{\text{max}}^{2}}{2\tau^{2}}\left(  \frac{d}
    {dt}\partial_{v}\mathcal{F}-\frac{V}{\lambda}\partial_{v}\mathcal{F}
    +\partial_{h}\mathcal{F}\right)  =0\,.
    \label{2.9}
\end{multline}
In the vicinity of the stationary point $( h_{V},V)$ its solution is
actually given by \textit{Ansatz}~(\ref{2.56}), which immediately
leads us to the following expressions for the amplitudes
\begin{subequations}\label{2.13}
\begin{eqnarray}
 h_{+}& = & \frac{\tau(v-V)-\zeta_{-}(h-h_{V})}{\zeta_{+}-\zeta_{-}}\,,
 \label{2.13a}\\
 h_{-}& = & \frac{\zeta_{+}(h-h_{V})-\tau(v-V)}{\zeta_{+}-\zeta_{-}}
 \label{2.13b}
\end{eqnarray}
\end{subequations}
and relates the car acceleration $a$ to the car velocity $v$ and the
headway distance $h$:
\begin{equation}
    a=-\frac{\left(\zeta_{+}+\zeta_{-}\right)}{\tau}\left[(v-V)-
    \frac{\zeta_{-}\zeta_{+}}{\left(\zeta_{+}+\zeta_{-}\right)}\frac{(h-h_{V})}{\tau}\right].
    \label{2.13aa}
\end{equation}
These results are analyzed individually.

\subsubsection{Optimal driving condition}

The stationary point $(h = h_V, v= V, a = 0)$ of equation~(\ref{2.9})
gives the headway $h_{V}$ which the driver chooses in order to follow
the lead car at speed $V$. In particular, the expression
\begin{equation}
 \partial_{h}\mathcal{F}-\frac{v}{\lambda}\partial_{v}\mathcal{F}=0
 \label{2.11}
\end{equation}
specifies the relationship between the values of the headway $h$ and
the car velocity $v$ for the stationary traffic flow imitated by the
given car following problem. Solving Eq.~(\ref{2.11}) for the car
velocity $v$ the optimal velocity approximation is obtained,
$v=\vartheta_{\text{opt}}(h)$. In particular, for the specific
form~(\ref{1.7}) of the cost function $\mathcal{F}(h,v)$ we
immediately get for $l\ll h\ll\lambda$
\begin{equation}
\vartheta_{\text{opt}}(h)=\vartheta_{\text{max}}\frac{h^{2}}{h^{2}+D^{2}}\,,
    \label{2.12}
\end{equation}
where the spatial scale $D$ is given by expression
\begin{equation}
D=\sqrt{\frac{\lambda l}{2}}\,,\quad l\ll D\ll \lambda\,.
    \label{2.51}
\end{equation}
%
Relation~(\ref{2.12}) or similar sigmoid functions are widely used in
current literature. In particular, for $l=1$~m and $\lambda = 300$~m
Exp.~(\ref{2.51}) gives the estimate $D\approx 12$~m typically
ascribed to the spatial scale $D$.

\subsubsection{Linear governing equation for the car following problem}

\begin{figure}
  \includegraphics[width=70mm]{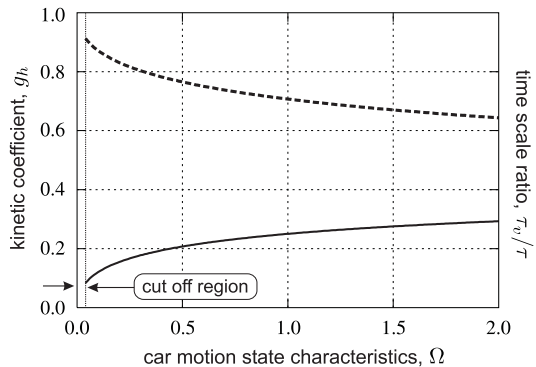}\\
  \caption{The kinetic coefficient $g_h$ and the ratio $\tau_v/\tau$ vs
    $\Omega$. 
    In drawing these curves
    Exp.~(\ref{2.804}) with $\Lambda = 1$ and $\phi = 0$ was used. The
    solid curve is $g_h(\Omega)$, the dashed curve is
    $\tau_v/\tau(\Omega)$.  \label{FBRDM}}
\end{figure}

By introducing the time scale $\tau_v = \tau/(\zeta_{+}+\zeta_{-})$
and the coefficient
\begin{equation}\label{gv}
    g_h = \frac{\zeta_{-}\zeta_{+}}{\left(\zeta_{+}+\zeta_{-}\right)^2}\,.
\end{equation}
the car dynamics equation~(\ref{2.13aa}) can be rewritten as
\begin{equation}\label{lge}
a=-\frac{1}{\tau_v}\left[(v-V)-g_h\frac{(h-h_{V})}{\tau_v}\right].
\end{equation}
Equation~(\ref{lge}) plays a significant role in models that describe
non-rational driver behavior \cite{BRDM},
%
%
where expression~(\ref{lge}) specifies an optimal acceleration that
could be chosen by rational drivers and the parameter $g_h$ was
introduced phenomenologically.
%
%
Here, $g_h$ can be computed using the cost function~(\ref{1.7}). It is
plotted as well as the ratio $\tau_v/\tau$ as a function of the
variable $\Omega$ (Fig.~\ref{FBRDM}). This representation is due to
the fact that the coefficient $\Lambda \sim 1$ (for the cost function
~(\ref{1.7}) $\Lambda \simeq 1$ for $l\ll h \ll \lambda$), the
coefficient $\phi \ll 1$ by virtue of the adopted
assumption~(\ref{2.sigma}), whereas the coefficient $\Omega$ varies
around unity. Below, the latter will be justified and the dependence
of the coefficient $\Omega$ on the car motion state will be analyzed.
As seen in Fig.~\ref{FBRDM} the time scales $\tau_v$ and $\tau$
practically coincide with each other and the parameter $g_h$ is a
small value.

\subsubsection{Relaxation curves vs system parameters}

Using again the cost function Eq.~(\ref{1.7}), and taking into account
$l\ll h_V\ll \lambda$ and $\Lambda \simeq 1$, the optimal velocity
dependence $\vartheta_{\text{opt}}(h_V)$ is given by Eq.~(\ref{2.12}).
The parameter $\Omega$ depends as follows on the car motion state
\begin{equation}\label{2.omega}
\Omega = 4\sigma \frac{\tau d\vartheta_{\text{opt}}(h)}{dh}\Big|_{h=h_V} =
    8\sigma^2\frac{\lambda D^2h_V}{(h^2_V + D^2)^2}\,.
\end{equation}
It attains the maximum
\begin{equation}\label{2.omegaMax}
\Omega_{\text{max}} = \frac{3\sqrt{3}\,(\tau\vartheta_{\text{max}})^2}
    {2D\lambda} =\frac{3\sqrt{3}\,\sigma^2}{2}\frac{\lambda}{D}
\end{equation}
at $h_{\Omega} = D/\sqrt{3}$. Using $\tau \sim 1$~s,
$\vartheta_{\text{max}}\sim 100$~km/h, $D\sim 15$~m, and $\lambda\sim
300$~m, respectively, $\Omega_{\text{max}}\sim 0.5$ is obtained. Of
course, this is not a precise numerical value, but in general
$\Omega_{\text{max}}\lesssim 1$ may be assumed.

\begin{figure}
  \includegraphics[width=70mm]{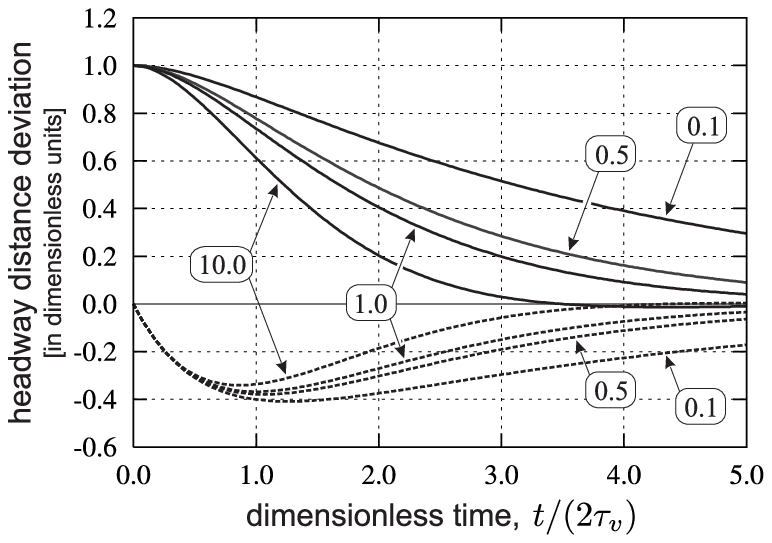}\\ \vspace{0.5\baselineskip}
  \includegraphics[width=70mm]{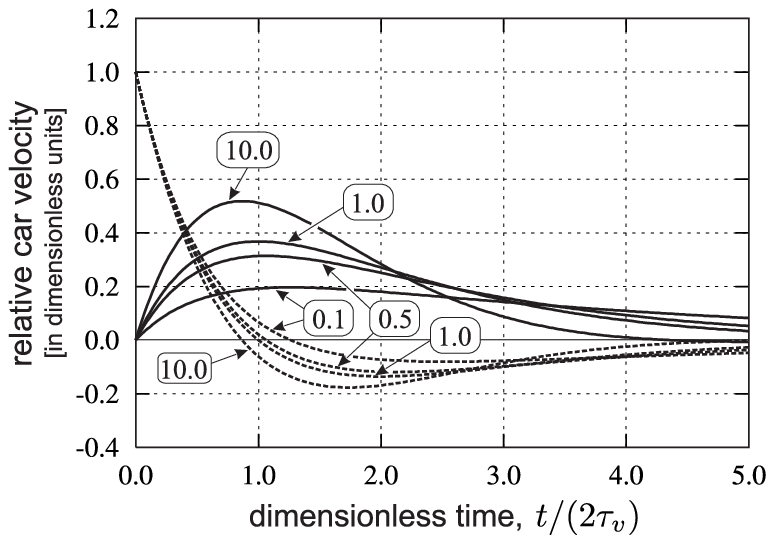} \\
  \caption{The relaxation curves of car dynamics for different values of
    the parameter $\Omega$ (labelled by numbers in squares).
    The solid curves display the relaxation caused by a perturbation
    in the headway, the dashed one shows the same for a velocity
    perturbation. The variations in the headway distance and relative
    velocity are measured either in units of the initial values
    $\delta h_0$ and $\delta v_0$ or $2\tau_v\delta v_0$ and $\delta
    h_0/(2\tau_v)$, respectively.
  \label{FRC}}
\end{figure}

The precise value of $\Omega_{\text{max}}$ matters. According to
Sec.~\ref{sec:ee} the car dynamics is characterized by two
eigenfunctions with eigenvalues given by Eq.~(\ref{2.804}). When
$\Omega < 1$, the relaxation is a pure fading process characterized by
time scales $\tau/\zeta_+$ and $\tau/\zeta_-$.  Otherwise, $\Omega >
1$, the relaxation process is characterized by the complex
eigenvalues, $\zeta_+$, $\zeta_-$. However in this region their
dependence on $\Omega$ is sufficiently weak,
$|\zeta_\pm|\sim \sqrt[4]{\Omega}$ for $\Omega\gg 1$
as follows from Exp.~(\ref{2.804}). Therefore the car relaxation
processes should be rather insensitive to the value of $\Omega$ as it
varies from unity to about ten.  Figure~\ref{FRC} demonstrates this
fact showing the relaxation curves induced by deviations from the
stationary values of the headway and the car velocity individually.
All the corresponding curves lie near each other, only for $\Omega =
0.1$ a larger deviation is visible. This is because for such values of
$\Omega$ the eigenvalues $\tau/\zeta_\pm$ differ essentially.
By contrast, the cost of acceleration, i.e., the cost of correcting
the car motion depends strongly on $\Omega$ because this component of
the cost function~(\ref{2.1}) varies linearly with $\Omega$ to the
first approximation. Since the maximum $\Omega_{\text{max}}$ of the
parameter $\Omega$ is attained at $h\sim D$ this means that the
greater the value of $\Omega$, the larger is the cost of the car
governing with respect to the cost of motion in its own accord.

\subsubsection*{Hypothesis about the value of $\vartheta_{\text{max}}$}

Keeping in mind the aforementioned speculations let us assume that the composed
parameter $\Omega_{\text{max}}$ takes a certain fixed numerical value,
$\Omega_{\text{max}} = \Omega_c \lesssim 1$. It leads immediately to a certain
relationship between the initial quantities of the model,
$\vartheta_{\text{max}}$, $\lambda$, $\tau$, $l$. In this collection only the
optimal velocity $ \vartheta_{\text{max}}$ on empty road is determined solely
by the driver behavior. The other parameters, namely, the size $\lambda$ of
recognition region, the characteristic time scale $\tau$ of the car ability to
accelerate or decelerate, and the characteristic headway $l$ that drivers can
control reliably and should be about the typical headway distance in dense jams
are determined by other mechanisms, e.g., car properties, driver physiology,
visual conditions and the like. At first glance such relations seem to be
impossible, because a driver chooses the optimal velocity on empty road
depending on the road conditions, speed regulations, weather, \textit{etc}.
Nevertheless, there is an additional factor affecting the car motion on empty
roads, the driver experience. If the driver preference for the empty road
velocity stems from here experience of driving in low density traffic flow then
such relations can exist.

In the given model this assumptions lead to the expression
\begin{equation}\label{2.vmax}
    \vartheta_{\text{max}} = \left[\frac{\sqrt{2}\,\Omega_c}{3\sqrt{3}}
    \right]^{\tfrac12}
    \frac{l^{\frac14}\lambda^{\frac34}}{\tau}
\end{equation}
Naturally, the specific form of relation~(\ref{2.vmax}) will change if, for
example, another \textit{Ansatz} of the cost function is used, however, its
qualitative features should be retained. These speculations, definitely, are
not enough to regard relations similar to formula~(\ref{2.vmax}) as an
established theoretical result. We have presented it for discussion only.
%

\subsubsection{Following-the-leader model vs optimal velocity model}

As discussed already there are two stimuli for a driver to change the
current motion state. One is the velocity difference $v-V$ between the
leading car and her car. The second is the difference between the
current speed $v$ and the optimal speed $\vartheta_{\text{opt}}(h)$ as
function of headway.  Models like (\ref{IModels}) take into account
both of them, leaving the determination of the weight coefficients to
appropriate empirical or experimental data. The results obtained so
far actually allow to predict these coefficients.

\begin{figure}
  \includegraphics[width=70mm]{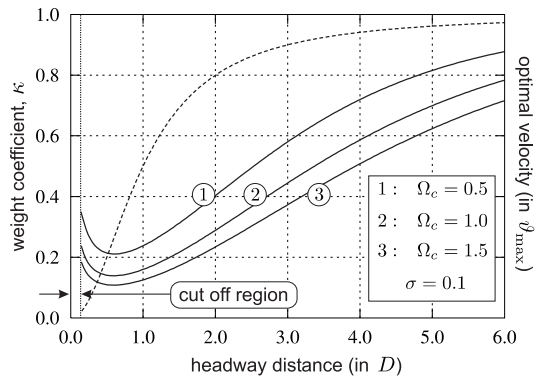}
  \caption{The weight coefficient $\kappa$ of the combined
    model~(\ref{CM}) vs the stationary headway $h_V$.
    For comparison, the dashed curve shows the optimal velocity,
    Eq.~(\ref{2.12}).\label{FKap}}
\end{figure}

This can be demonstrated with the combined model~(\ref{CM}).
Linearizing Eq.~(\ref{CM}) around the stationary point $(h_V,V)$ and
comparing the result with Eq.~(\ref{2.13aa}) we get
\begin{align}
\kappa &  =\frac{\zeta_{-}\zeta_{+}}{\left(  \zeta_{+}+\zeta_{-}\right) }\left[
    \frac{\tau d\vartheta_{\text{opt}}}{dh}\Big|_{h=h_{V}}\right]  ^{-1}.
    \label{2.cfp4}
\end{align}
This expression gives $\kappa$ as function of the car motion state.
Figure~\ref{FKap} plots this dependence for the cost
function~(\ref{1.7}). As seen in Fig.~\ref{FKap} the weight
coefficient $\kappa$ is small for all interesting values of the
headway distance in the car following regime. Only for small headways
$h_V\sim l$ (dense jams) or for large headways $h_V$ exceeding $D$
substantially (free flow) $\kappa$ approaches unity.

\subsubsection{Different types of the car dynamics\label{sec:TCMD}}

\begin{figure}
  \includegraphics[width=80mm]{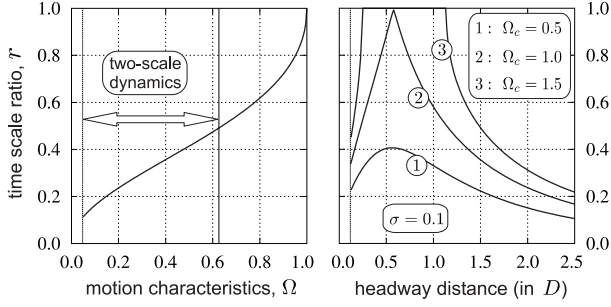}
  \caption{The ratio of time scales $r = {|\zeta_-|}/{|\zeta_+|}$ of the
    corresponding eigenfunctions describing the car motion relaxation
    as a function of $\Omega$ or the headway $h_V$ when $\Omega_c$ is
    fixed. Computations in this figure used cost
    function~(\ref{1.7}) and $\phi=0$.\label{FRO}}
\end{figure}

As the headway distance $h_V$ varies in the interval $l\ll h_V\ll
\lambda$ the coefficient $\Omega(h_V)$ changes essentially, leading to
qualitatively different car dynamics. As mentioned before, when
$\Omega(h_V)>1$ the relaxation exhibits damped oscillations around the
stationary point $(h_V,V)$.  This is due to the eigenvalues
$\{\zeta_+, \zeta_-\}$ (see Exp.~(\ref{2.804})) having non-zero
imaginary parts. Since for $\Omega(h_V)\gtrsim 1$ their real and
imaginary parts are of the same order the car motion relaxation is
characterized by one time scale about one (in units of $\tau$). This
is true also for $\Omega(h_V)\lesssim 1$ because $\zeta_+ \approx
\zeta_-$ there, although the dynamics is a pure fading process now.
For still smaller values of $\Omega(h_V)\approx 0.6$ the ratio
$r=|\zeta_-|/|\zeta_+|$ of these eigenvalues becomes smaller than one
half, which may be defined as the two-scale regime of the dynamics,
see Fig.~\ref{FRO}.

The change between the two-scale and the one-scale regime corresponds
to a value of $h_V \approx 1$--1.5~$D$. Below this value, the
dynamical behavior is one-scale, above it is two-scale and will be
called ``fast-and-slow'' in the following.

Concerning with the ``fast-and-slow'' dynamics the velocity relaxation
and the headway relaxation can be analyzed individually. According to
Exps~(\ref{2.13}) the initial difference $h-h_V$ contributes mainly to
the eigenfunction with the eigenvalue $\zeta_-$. Thereby, the velocity
difference $v-V$ contributes mostly to the amplitude $h_+$. The
amplitude $h_-$ also contains the term of the same magnitude, however,
the time scale $\tau/\zeta_-$ on which the corresponding eigenfunction
varies is much bigger than the time scale $\tau/\zeta_+$ of the other
eigenfunction. So, the velocity relaxation falls on the first
eigenfunction. Thus, the velocity difference $v-V$ disappears
practically completely during the time $\tau/\zeta_+$, which forms the
``fast'' stage of the car relaxation. At the next ``slow'' stage of
duration $\tau/\zeta_+$ the headway deviation from the equilibrium
value $h_V$ disappears. To summarize, the ``fast-and-slow'' car
dynamics is a two-stage process where the velocity difference between
the cars is eliminated first. Later on, the headway is optimized.
While this is being done, the resulting velocity difference is not
essential and the driver can govern the car motion without the
necessity of responding fast.

\begin{figure}
  \includegraphics[width=70mm]{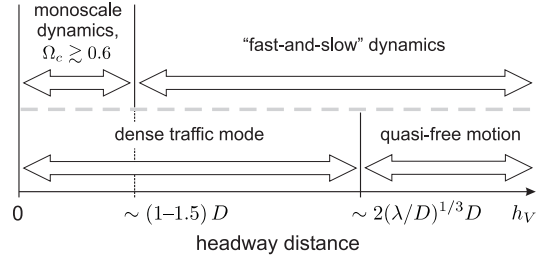}
  \caption{Possible types of the car motion state classified using different
    properties of the car dynamics and their mutual arrangement
    depending on $h_V$. \label{FOL}}
\end{figure}

Another important characteristics, separating dense traffic and
quasi-free flow can be derived as follows. The solution of the
eigenvalue equation~(\ref{2.800}) depend actually on two values,
$\phi$ and $\Omega$, since $\Lambda \simeq 1$ for the cost
functional~(\ref{1.7}). The parameter $\phi < \sigma\ll 1$ by virtue
of the adopted assumption~(\ref{2.sigma}). The maximum of $\Omega$
attained at $h_V=D/\sqrt{3}$ is much larger than $\sigma^2$ as it
results from Exp.~(\ref{2.omegaMax}). However, as the headway distance
increases the value $\Omega(h_V)$ decreases as $h_V^{-3}$ (see
Exp.~(\ref{2.omega})) whereas $\phi\to\sigma$. So there is a value
$h_c$ of the headway distance at which both these terms contribute to
the eigenvalues to the same extent. To find $h_c$ consider $h\gg D$
where $\Omega(h_V)\ll 1$ and $\phi\approx \sigma$, thereby, $\zeta_+
\simeq 1$ and
\begin{equation}\label{2.add1}
    \zeta_- \simeq \frac{\Omega(h_V)}{2\big(\sigma + \sqrt{\sigma^2 +
    \Omega(h_V)}\,\big)}
\end{equation}
by virtue of formula~(\ref{2.804}). So, $h_c$ is specified by
\begin{equation}\label{2.add2}
\Omega(h_c) = \sigma^2\quad \Rightarrow \quad
    h_c = 2\left(\frac{\lambda}{D}\right)^{\tfrac13}D\,.
\end{equation}
%
The value $h_c$ divides the headway distance region into two parts,
see Fig.~\ref{FOL}. When $h_V\gtrsim h_c$ the velocity of the
following car is close to $\vartheta_{\text{max}}$, so this type of
car motion will be referred to as the quasi-free flow. In this case
the eigenvalue equation~(\ref{2.800}) cannot be simplified, so to
describe the quasi-free flow the cost functional~(\ref{2.7}) only in
its full form may be used. In the opposite case, $h\ll h_c$, which
will be called dense traffic mode, the term $\phi$ is ignorable,
reducing the eigenvalue equation~(\ref{2.800}) to
\begin{equation}\label{2.add3}
\zeta^{4}-\Lambda\zeta^2 +\frac{1}{4}\Omega=0\,.
\end{equation}
Moreover, in the latter case the description of car dynamics can be
reduced to the standard form of classical mechanics, enabling us to
analyze the nonlinear stage of the car dynamics.

\section{Nonlinear car dynamics\label{sec:ncd}}

\subsection{Effective cost functional for dense traffic flow}

%
%
In the dense traffic limit $h\ll h_c$, the variational principle based
on optimizing the cost functional~(\ref{2.7}) can be simplified
essentially. In the given limit the characteristic time scales of the
car dynamics are $\tau$ and $\tau/\sqrt{\Omega(h_V)}$, with both of
them being small in comparison with $\lambda/\vartheta_{\text{max}}$.
The latter follows from the adopted assumption~(\ref{2.sigma}) and the
condition $\Omega(h_V)\gg \sigma^2$ for $h\ll h_c$.

In this case, as shown in Appendix~\ref{app:B}, the cost
functional~(\ref{2.7}) can be replaced by the following effective
functional whose integrand does not contain a time dependent factor
\begin{equation}
    \mathcal{L}\left\{h(t)\right\}  =\int_{t}^{\infty}\mathcal{F}
    _{\text{eff}}\left(h,v,a\mid V\right)dt'\,.
    \label{10.2}
\end{equation}
where the integrand, which will be called the Lagrangian of the car
dynamics is given by
\begin{multline}
    \mathcal{F}_{\text{eff}}\left(h,v,a\mid V\right)  =
    \mathcal{F}^d\left( h,v,a\right)\\
    -\left( \frac{V}{\lambda}\,h + v\right)
    \left.\partial_{v}\mathcal{F}\right\vert_{V,h_{V}}\,.
    \label{10.3a}
\end{multline}
Interestingly, $\mathcal{F}_{\text{eff}}\left( h,v,a\mid V\right)$
contains the lead car velocity $V$ as a parameter and attains its
extremal value with respect to $h$ at the point $h_{V}$ corresponding
to the optimal driving with the velocity $V$. The term
$v\,\partial_{v}\mathcal{F}\vert_{V,h_{V}}$ has been introduced for
the sake of convenience only, it does not affect the extremal equation
but enables the Lagrangian to attain a minimum with respect to the car
velocity $v$ at the stationary point $(h_V,V)$.

Both functionals~(\ref{2.7}) and (\ref{10.2}) possess the same
extremals to the first order in the small parameter
$\sigma/\sqrt{\Omega(h_V)}$. In particular, the extremals of
functional~(\ref{10.2}) meet the equation
\begin{equation}
    \frac{2\tau^{2}}{\vartheta_{\text{max}}^{2}}\frac{d^{2}a}{dt^{2}}-\frac{d}
    {dt}\partial_{v}\mathcal{F}-
    \partial_{h}\left(\mathcal{F}-\frac{V}{\lambda }\left.
    \partial_{v}\mathcal{F}\right\vert _{V,h_{V}}h\right)  =0\,.
    \label{10.0}
\end{equation}
It corresponds directly to the initial full equation~(\ref{2.9}) where
the second and third terms in the former parentheses are ignored
whereas the second term in the latter parentheses is replaced by its
value taken at the stationary point. It is justified because these
terms are due to variations of the time dependent cofactor.
%
%

Keeping in mind its following applications we rewrite Lagrangian
$\mathcal{F}_{\text{eff}}\left( h,v,a\mid V\right)$ also in the form
\begin{equation}\label{10.3b}
    \mathcal{F}_{\text{eff}}\left(h,v,a\mid V\right) =
    \frac{\tau^2a^2}{\vartheta^2_{\text{max}}} + \mathcal{F}_0(v) +
    \mathcal{F}_{\text{int}}(h,v\mid V)\,.
\end{equation}
In particular, for the cost function~(\ref{1.7}) we get
\begin{align}\label{10.301}
    \mathcal{F}_0(v\mid V) & =\frac{(v-V)^2}{\vartheta_{\text{max}}^2}+
    \left(1- \frac{V^2}{\vartheta_{\text{max}}^2}\right)\,,\\
    \mathcal{F}_{\text{int}}(h,v\mid V) & =
    \frac{\Omega\, h_V^2}{4\vartheta_{\text{max}}^2\tau^2}
    \left(\frac{v^2}{V^2}\frac{h_V}{h}+\frac{h}{h_V}\right).
    \label{10.302}
\end{align}
In obtaining formula~(\ref{10.302}) we have taken into account the
expression~(\ref{2.12}) relating the velocity $V$ to the headway
distance $h_V$, expression~(\ref{2.omega}), and omitted some
insignificant terms. The latter term in expression~(\ref{10.301}) can
be also omitted because it has no effect on the extremal equation.

The most essential feature of the effective cost
functional~(\ref{10.2}) is the absence of a time dependent cofactor.
This enables us to reformulate the car dynamics in terms of autonomous
Hamiltonian equations.

\subsection{Hamiltonian description of car dynamics}

Finding the extremals of functional~(\ref{10.2}) can be done as
follows. The phase space $\{h,v,a\}$ is expanded to
$\{h,\dot{h},v,\dot{v}\}$ by adding the relationship between the
headway distance $h$ and the car velocity $v$ as an additional
constraint:
\begin{gather}
 \text{minimize}\quad\int_{0}^{\infty}\mathcal{F}_{\text{eff}}
 \left(h,v,\dot{v}\right)dt
 \label{app.4}
\intertext{subject to the dynamical equation}
 \dot{h}= V-v \label{app.5}
\end{gather}
and the initial and final conditions,
\begin{align}
 \label{app.6a}
 h(0)& =h_{0}\,,& v(0) &=v_{0}\,, \\
 h(\infty)& =h_{V}\,,& v(\infty)&=V\,.
 \label{app.6b}
\end{align}
For the sake of simplicity the parameter $V$ has been omitted in the
list of variables of $\mathcal{F}_{\text{eff}}$, the current time is
set to zero.
By using a Lagrange multiplier, the problem~(\ref{app.4}) is rewritten
in the standard form.  Namely, the extremals of
\begin{equation}
 \int_{0}^{\infty}\left[\mathcal{F}_{\text{eff}}\left(h, v,\dot{v}\right)
 + p\,(\dot{h}+ v- V)\right]dt
 \label{app.7}
\end{equation}
are sought. They are defined on the extended set of variables
$\left\{h(t),v(t),p(t)\right\}$ and subject to
conditions~(\ref{app.6a}), (\ref{app.6b}). The extremals of
functional~(\ref{app.7}) obey the classical Lagrange equations.

In constructing the Hamiltonian $\mathcal{H}(h,v,p,q)$ that produces
the same extremal equations the Pontryagin technique is used
\cite{Pontr}. Introducing the new variable $q$
\begin{equation}\label{app.q}
    q = \frac{\partial\mathcal{F}_{\text{eff}}(h,v,\dot{v})}{\partial\dot{v}}
\end{equation}
and solving for $\dot{v}$, i.e., finding $\dot{v}=\dot{v}(h,v,q)$, the
desired Hamiltonian is written as
\begin{equation}\label{app.H}
    \mathcal{H}(h,v,p,q) = q\dot{v} - \mathcal{F}_{\text{eff}}(h,v,\dot{v}) - p(v-V)\,.
\end{equation}
It can be demonstrated directly that the desired extremals meet the
standard Hamiltonian system of equations
\begin{align}
    \label{app.H1}
    \dot{h} & = \partial_p \mathcal{H}\,, &
    \dot{v} & = \partial_q \mathcal{H}\,, \\
    \label{app.H2}
    \dot{p} & = -\partial_h \mathcal{H}\,, &
    \dot{q} & = -\partial_v \mathcal{H}\,.
\end{align}

Again, the optimization of functional~(\ref{app.4}) actually leads to
a boundary value problem because the extremal is specified by both the
initial and final conditions~(\ref{app.6a}), (\ref{app.6b}).  The
Hamiltonian approach~(\ref{app.H1})--(\ref{app.H2}) also shares this
property. Namely, the initial values of the variables $h$ and $v$ are
given by conditions~(\ref{app.6a}), whereas the initial values $p_0$
and $q_0$ of the quasi-momenta $p$ and $q$ should be chosen such that
the system tend to the stationary point $(h_V,V)$ as $t\to\infty$. It
is the same situation that we met in dealing with equation~(\ref{2.9})

However, the Hamiltonian description has a certain advantage. First,
the Hamiltonian itself is conserved,
\begin{equation}\label{app.H3}
    \frac{d\mathcal{H}(h,v,p,q)}{dt} = 0\,.
\end{equation}
Second, an additional autonomous first integral of the
system~(\ref{app.H1}), (\ref{app.H2}), can be found, at least, for a
certain stage of the car dynamics. Fortunately, this is the case for
the ``fast-and-slow'' car dynamics analyzed below.

Note, that this Hamiltonian description of the car motion relaxation
towards the stationary state does not contradict the conservation of
phase volume being the general property of Hamiltonian systems. The
matter is that the relaxation process is described by only
\textit{one} path leading to the stationary point of the saddle type.
Other possible paths are not considered.

\subsubsection*{Physical meaning of the Hamiltonian variables}

The two Eqns~(\ref{app.H1}) can be understood readily. The equation
for $\dot h$ is just $\dot h = V -v$, while the equation for $\dot q$,
or, what is the same, expression~(\ref{app.q}) shows that the
quasi-momentum $q$ is proportional to the car acceleration
$a=\dot{v}$:
\begin{equation}\label{10.100}
    q = \frac{2\tau^2}{\vartheta^2_{\text{eff}}}\,a\,.
\end{equation}

The physical meaning of the quasi-momentum $p$ is more complex. To
clarify it, rewrite the equation line~(\ref{app.H2}) in terms
of partial derivatives of the Lagrangian
$\mathcal{F}_{\text{eff}}(h,v,\dot{v})$. Namely, the definitions of
the quasi-momentum $q$~(\ref{app.q}) and the Hamiltonian~(\ref{app.H})
enables us to represent these equations as
\begin{align}
    \label{10.101}
    \dot{q} & = \partial_v \mathcal{F}_{\text{eff}} + p\,,\\
    \label{10.102}
    \dot{p} &= \partial_h \mathcal{F}_{\text{eff}}\,.
\end{align}
In particular, expressions~(\ref{10.100})--(\ref{10.102}) immediately
lead, as it must, to equation~(\ref{10.0}).

Expression~(\ref{10.101}) demonstrates that the quasi-momentum $p$
includes the rate $\dot{a}$ of acceleration variations. In this way
the variable $\dot{a}$ implicitly enters the Hamiltonian
$\mathcal{H}(h,v,p,q)$ specifying a hyper-surface in the
four-dimensional phase space $\{h,v,a,\dot{a}\}$. Even more, by
ignoring the dependence of $\mathcal{F}_{\text{eff}}(h,v,\dot{v})$ on
the headway distance $h$, the velocity relaxation towards the
stationary value $V$ is governed by the equation
\begin{equation*}
    \dot{q}  = \partial_v \mathcal{F}_{\text{eff}}\,.
\end{equation*}
It results from the effective cost functional~(\ref{10.2}) where the
car velocity $v$ is treated as a primary argument. As will be seen
below, there is a stage of the car dynamics where this assumption is
justified and the quasi-momentum $p = 0$ is conserved. Only the
dependence of the Lagrangian $\mathcal{F}_{\text{eff}}(h,v,\dot{v})$
on the headway distance $h$, see Eq.~(\ref{10.102}), causes variations
in the quasi-momentum $p$. Since the velocity control is of primary
importance in the driving strategy, the variable $p$ can be regarded
as a certain measure of the necessity to control also the headway
distance when eliminating the velocity difference.

\subsection{``Fast-and-slow'' dynamics}

For $\Omega\ll 1$ the last term on the right-hand side of
Exp.~(\ref{10.3b}) is small. Since the headway $h$ enters the
Hamiltonian~(\ref{app.H}) exactly via this term time variations in the
quasi-momentum $p$ are retarded by virtue of the first equation at
line~(\ref{app.H2}). The same follows from Eq.~(\ref{10.102}). The
other variables $h$, $v$, $q$ can vary substantially on scales
independent of the value $\Omega\ll 1$. Then, the car dynamics can
exhibit multi-scale relaxation. In particular, Sec.~\ref{sec:TCMD}
demonstrated this fact, namely, it has been shown that in the given
limit the system dynamics comprises the fast and slow stages. Exactly
the time scale of the latter stages depends on the value $\Omega\ll
1$. During the former stage the quasi-momentum $p$ should be
practically constant, i.e., it is approximately a first integral of
the system~(\ref{app.H1}), (\ref{app.H2}). This property, together
with the conservation of Hamiltonian~(\ref{app.H}), allows the
integration of the system of equations~(\ref{app.H1}), (\ref{app.H2}).
During the following slow stage a quasi-stationary approximation can
be used to analyze the system evolution.

The results below are exemplified with the cost function~(\ref{1.7}).
They can be easily generalized to other cost functions.

\subsubsection{Fast stage}

At the zeroth approximation in the small parameter $\Omega$ the
quasi-momentum $p$ is a constant. During the fast stage mainly the
velocity deference $v-V$ is eliminated, so Eq.~(\ref{10.101}) formally
describes the relaxation process of the car velocity $v$ to the
stationary value $V$. Therefore, by virtue of Exp.~(\ref{10.301}), the
quasi-momentum $p$ takes zero value, $p=0$. In this limit the
Lagrangian component $\mathcal{F}_{\text{int}}(h,v\mid V)$ as well as
the additive constant of the component $\mathcal{F}_0(v\mid V)$ can be
omitted. So, the Hamiltonian~(\ref{app.H}) becomes
\begin{gather}
    \nonumber
    \mathcal{H}(v,a)  = \mathcal{H}_a(a)-\mathcal{H}_v(v)\,,\\
\intertext{where}\label{10.103}
    \mathcal{H}_a(a)  = \frac{\tau^2a^2}{\vartheta_{\text{max}}^2}\,, \qquad
    \mathcal{H}_v(v)  = \frac{(v-V)^2}{\vartheta_{\text{max}}^2}\,,
\end{gather}
and the non-canonical variables $\{h,v,a\}$ are used.

Since the Hamiltonian $\mathcal{H}(v,a)$ is conserved and the fast
stage describes the velocity relaxation to $V$ the car dynamics obeys
the equation
\begin{equation}\label{fs.1}
    \mathcal{H}_a(a)  =\mathcal{H}_v(v)\,.
\end{equation}
This immediately gives the relationship between the acceleration $a$
and the velocity $v$
\begin{equation}\label{nad12}
    a = -\frac1{\tau}(v-V)\,.
\end{equation}
The sign in the latter equation has been chosen so to allow for the
system relaxation.

Equation~(\ref{fs.1}) is, in fact, of the general form and the linear
form of Eq.~(\ref{nad12}) is due to the adopted quadratic
\textit{Ansatz} of the cost function but not a consequence of
linearization.  Besides, equality~(\ref{fs.1}) can be read as follows.
The comparison of the Hamiltonian parts $\mathcal{H}_a(a)$ and
$\mathcal{H}_v(v)$ with the corresponding components of
Lagrangian~(\ref{10.3b}) demonstrates that the function
$\mathcal{H}_a(a)$ actually measures the cost of the car acceleration
and the function $\mathcal{H}_v(v)$ does the same with respect to the
car motion relative to the optimal driving conditions. Thereby, during
the fast stage the driver corrects the car dynamics so that the cost
of acceleration be equal to the cost of current motion measured
relative to the stationary conditions.

\begin{figure}
  \includegraphics{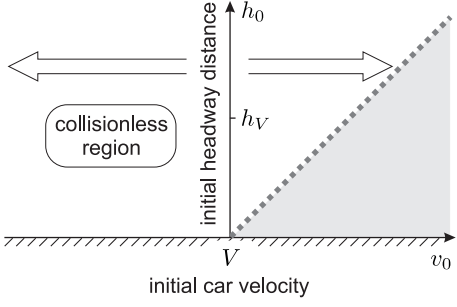}
  \caption{Phase region of the initial values of the headway distance
    $h_0$ and the car velocity $v_0$ where the car dynamics is
    collisionfree (based on equation~(\ref{nad13})).\label{FCol}}
\end{figure}

The headway distance $h$ does not enter explicitly the governing
equation of the fast stage. However, it varies during the fast stage
and finally attains a value $h_*$ differing from the initial value
$h_0$. In the adopted simple approximation of the cost function
Eq.~(\ref{nad12}) enables to estimate easily the value $h_*$. Namely,
the direct integration of Eq.~(\ref{nad12}) yields the relationship
\begin{equation}\label{nad13}
    h_* = h_0 - \tau(v_0 - V)\,.
\end{equation}
Applying to formula~(\ref{nad1010}) it can be seen that
expression~(\ref{nad13}) holds until the headway distance $h$ becomes
too small and it is impossible to ignore the effect of the term
$\mathcal{F}_{\text{int}}(h,v)$, i.e., when
\begin{equation}\label{nad14}
    h\lesssim \sqrt[3]{\Omega}\, h_V \ll h_V\,.
\end{equation}
The car dynamics in the region of small values of the headway distance
when the probability of collision is high is worth of an individual
consideration. Here we only touch on this problem by assuming the
collision to happen when the value $h_*$ given by Exp.~(\ref{nad13})
becomes equal to zero. This assumption is justified as a rough
approximation due to estimate~(\ref{nad14}), see Fig.~\ref{FCol}. The
boundary of the collisionless region can be shifted to the right
substantially by dropping the assumed quadratic dependence of the cost
function on acceleration. This will be done elsewhere.

\subsubsection{Slow stage}

When the system attains a quasi-equilibrium with respect to the car
velocity $v$ its further evolution is due to the direct dependence of
the Hamiltonian $\mathcal{H}(h,v,q,p)$ on the headway distance $h$. It
enters the Hamiltonian via $\mathcal{F}_{\text{int}}(h,v\mid V)$. As a
result, the velocity difference $v-V$ and the acceleration $a$ should
be small. Keeping this in mind, Eq.~(\ref{10.101}) can be solved for
$p$. Substitution of the obtained expression into Eq.~(\ref{10.102}),
and the further linearization with respect to the variable $v-V$ and
the component $\mathcal{F}_{\text{int}}$ results in
\begin{equation}\label{st.1}
        \ddot{q}  - a\left. \partial_v^2 \mathcal{F}_0\right|_{v=V} =\partial_h
        \mathcal{F}_{\text{int}}\,.
\end{equation}
If $\tau_s$ is the characteristic time scale of the slow stage then
\begin{equation*}
    (v-V)\sim \frac{h-h_V}{\tau_s}\,,\quad
    a\sim\frac{h-h_V}{\tau^2_s}\,,\quad
    q\propto \dot{a}\sim\frac{h-h_V}{\tau^3_s}\,.
\end{equation*}
These estimates together with Eq.~(\ref{st.1}) lead to
%
\begin{equation}\label{st.2}
    \tau_s \sim \Omega^{-1/2}\tau.
\end{equation}
In particular, the rate $\dot{q}$ of time variations in the
quasi-momentum $q$ scales as $\dot{q}\propto \Omega^{2}$ for
$\Omega\to 0$. Since the Lagrangian $\mathcal{F}_{\text{eff}}$ has a
minimum at $h=h_V$ the quasi-momentum $p$ can be estimated as
\begin{equation*}
    p\sim \tau_s \partial_h \mathcal{F}_{\text{int}}\propto \Omega^{1/2}(h-h_V)
\end{equation*}
by virtue of Eq.~(\ref{10.102}). In the limit $\Omega\ll 1$ the term
$\dot{q}$ can be ignored in comparison with the quasi-momentum $p$.
Then, Eq.~(\ref{10.101}) immediately leads to the relation
\begin{equation}\label{st.3}
    p = -\partial_v \mathcal{F}_{\text{eff}}\,,
\end{equation}
which is no more than the standard expression for the momentum of a
system described by the Lagrangian $\mathcal{F}_{\text{eff}}(h,v,0)$
with $v = V -\dot{h}$.

By the same reasons all the terms in Hamiltonian~(\ref{app.H})
containing the quasi-momentum $q$ may be omitted. Then the
substitution of (\ref{st.3}) into (\ref{app.H}) yields the Hamiltonian
of the slow stage
\begin{align}
    \label{st.4}
    \mathcal{H}(h,v) & = \mathcal{H}_v(v) -\mathcal{H}_h(h)\,,\\
    \intertext{where}\nonumber
    \mathcal{H}_h(h) & = \mathcal{F}_{\text{int}}(h,V\mid V)-
    \mathcal{F}_{\text{int}}(h_V,V\mid V)\\
    \label{st.5}
     & = \frac{\Omega\, h_V^2}{4\vartheta_{\text{max}}^2\tau^2}
    \,\frac{(h-h_V)^2}{h_V h}\,.
\end{align}
In the Lagrangian component $\mathcal{F}_{\text{int}}(h,v\mid V)$ the car
velocity $v$ has been replaced by its stationary value $V$, and some negligible
terms have been omitted. Then the conservation of the Hamiltonian
$\mathcal{H}(h,v)$ during the system dynamics gets the form
\begin{align}
    \label{st.6}
    \mathcal{H}_v(v) & =\mathcal{H}_h(h)\\
\intertext{which leads to the following governing equation}
    \label{st.7}
    v-V &=\frac{\sqrt{\Omega}}{2\tau}\, (h-h_{V}) \sqrt{\frac{h_{V}}{h}}\,.
\end{align}
The conservation law~(\ref{st.6}) can be understood in analogy to that
for the fast stage. However, during the slow stage of the car dynamics
the car velocity plays the role of the control parameter. The driver
changes the speed to correct the headway distance. Again the cost of
deviation of the car velocity from the stationary value is chosen to
be equal to the cost of the car motion measured relative to the
optimal conditions.

It should be pointed out that the slow stage of the car dynamics is
governed by the conservation law, Exp.~(\ref{st.6}), which contains
only the headway distance $h$ and the velocity $v$. At the first
approximation, the acceleration $a$ does not enter at all. In this
meaning, the slow stage is similar to other physical systems.
However, the stationary point of the car dynamics is a saddle point
rather than a minimum.

Up to now, the two different stages of car-following discussed already
in Sec.~\ref{sec:TCMD} have been identified with two different
conservation laws.  Both of them can be unified into just another
effective conservation law interpolating Exps.~(\ref{fs.1}) and
(\ref{st.6}):
\begin{equation}\label{app.final}
    \mathcal{H}_{v}(v) = \mathcal{H}_a(a)+\mathcal{H}_h(h)\,.
\end{equation}
Within the same accuracy it is possible to interpolate directly
Eqs.~(\ref{nad12}) and (\ref{st.7}), leading to
\begin{equation}
    a = -\frac{1}{\tau}\Big[(v-V)-\frac{\sqrt{\Omega}}{2\tau}
    \Big(\frac{h_{V}}{h}\Big)^{\tfrac12}
    (h-h_{V})\Big].
    \label{nad1010}
\end{equation}

Besides, the proposed interpretations of the conservation
laws~(\ref{fs.1}) and (\ref{st.6}) enable us to formulate a
generalized principle of adequate control. It declares that the
effective cost of correcting the car motion via changing the car
acceleration (fast stage) or the car velocity (slow stage) is equal to
the cost of the current state of car motion measured with respect to
the optimal driving conditions.

\section{Conclusion and discussion\label{sec:CR}}

A variational approach to the description of car dynamics has been
developed in case of a car following a lead car moving at a constant
speed. To derive governing equations for the following car motion the
driver preference has been used to construct a cost functional.  Its
extremals specify the optimal paths of the further motion for the
following car. Applying to the general properties of the driver
behavior we analyzed the basic properties of the cost function and
proposed a simple parabolic \textit{Ansatz}, which, nevertheless,
catches typical features of traffic properties.

The concept of a rational driver is formulated. It comprises the
assumptions that the driver follows the optimal paths and corrects the
car motion continuously. In this case the optimal path is a Nash
equilibrium of the system, which is defined as follows.  If the driver
has chosen an optimal path at a certain moment of time then no further
correction of the car motion is necessary because it leads to the same
result.
As the consequence of the Nash equilibrium the extremals of the cost
functional specify the real dynamics of cars with rational drivers
although originally they determine only the imaginary paths in the
driver's mind when planning the further motion. In this way we
obtained several results. The optimal velocity approximation has been
derived. The weight coefficient entering a car following model
combining the ``following-the-leader'' model and the optimal velocity
model has been found depending on the headway distance.
As an important result it has been shown that the car dynamics can be
categorized under different types according to its properties. First,
we have shown that there can be two types of the car relaxation
towards the stationary motion, the mono-scale dynamics and the
``fast-and-slow'' dynamics. Second, we singled out the quasi-free
motion and the dense traffic mode.

The variational technique for the latter mode can be simplified
essentially.  Namely, it is possible to reformulate the cost
functional so that it does not contain a time dependent cofactor. As a
result, the autonomous Hamiltonian description for the car dynamics
has been constructed. For the dense traffic mode the ``fast-and-slow''
dynamics has been analyzed for the nonlinear stage. In particular,
different conservation laws for this stages have been found.  A
generalized principle of adequate car motion control has been
proposed.

We assumed that the lead car moves at a fixed speed. If it is not the
case two different situations should be singled out. For perfect
drivers who can predict rigorously the motion of the car ahead the
main results still hold. Otherwise the cost function will depend not
only on the current time but also, what is crucial, on the time when
the driver has started to evaluate the further car dynamics. This
disturbs the Nash equilibrium and the car correction should be carried
out continuously. Briefly this problem was studied in~\cite{we1} but
it actually requires a more detailed investigation.

\subsection*{Beyond the rationality}

Real drivers have certain limitations. They are not capable of finding
the optimal path precisely and they cannot correct the car motion
continuously. So, the concept of rational driver behavior is just the
first approximation of the real situation and deviations from this
perfect behavior should be analyzed consistently. A first step towards
this problem can be found in Ref.~\cite{BRDM}. Here we justify some
assumptions adopted there and substantiated them in a different way.

Let us again concern the driver evaluation of the car motion quality.
For a given path of the car's further motion
$\left\{\mathfrak{h}(\mathfrak{t},t)\right\}$ the cost
functional~(\ref{2.7}) is written as
\begin{multline}\label{br.1}
    \mathcal{L}\{\mathfrak{h}\} =
    \int_{t}^{\infty}d\mathfrak{t}\,e^{-\tfrac{V}{\lambda}(\mathfrak{t}-t)}
    \\ \times
    \left[\frac{\tau^{2}\mathfrak{a}^{2}}{\vartheta_{\text{max}}^{2}}
    +\left(  1-\frac{\mathfrak{u}}{\vartheta_{\text{max}}}\right)^{2}
    +\frac{\mathfrak{u}^{2}}{\vartheta_{\text{max}}^{2}}
    \frac{l}{\mathfrak{h}}\right]  \,.
\end{multline}
Expression~(\ref{br.1}) can be represented also in the form
\begin{align}
    \mathcal{L} \{\mathfrak{h}\} &
    =\int_{t}^{\infty}d\mathfrak{t}\,e^{-\tfrac{V}{\lambda}(\mathfrak{t}-t)}
    \nonumber\\
    & \times \left[\frac{\tau^{2}\mathfrak{a}^{2}}{\vartheta_{\text{max}}^{2}}
    + \frac{(\mathfrak{u}-V)^{2}}{\vartheta_{\text{max}}^{2}}
    +\frac{V^{2}}{\vartheta_{\text{max}}^{2}}\frac{l\delta^{2}}
    {h_{V}^{2}\left(h_{V}+\delta\right)}\right]
    \nonumber\\
    &  + \frac{2V}{\vartheta_{\text{max}}^{2}}(h-h_{V})+\mathcal{L}_{0}\, ,
    \label{br.2}
\end{align}
where $\delta(\mathfrak{t},t)$ describes the deviation of the given
path from the stationary car motion trajectory,
\begin{equation*}
    \mathfrak{h}(\mathfrak{t},t)=h_{V}+\delta(\mathfrak{t},t)\,.
\end{equation*}
Here, $\mathcal{L}_{0}$ is the cost of stationary car motion,
\begin{equation}
    \mathcal{L}_{0}=\frac{\lambda}{V}\left(\frac{V}{\vartheta_{\text{max}}
    }-1\right) ^{2}.
    \label{br.3}
\end{equation}
In addition, when deriving expression~(\ref{br.1}) we have used
\begin{equation*}
    \frac{l}{h}=\frac{l}{h_{V}}-\frac{l\delta}{h_{V}^{2}}+\frac{l\delta^{2}}
    {h_{V}^{2}\left(  h_{V}+\delta\right)  } ,
\end{equation*}
integrated various fragments of Eq.~(\ref{br.1}) by parts, assuming
$l\ll\mathfrak{h\ll}\lambda$, and dropped terms like $l/\mathfrak{h}$
where ever possible.

When the driver plan her further motion the current values of the
headway $h$ and the velocity $v$ are regarded as the initial
conditions. So, when choosing the optimal path she can minimize only
the first term $\mathcal{L}_{c}\left\{\mathfrak{h}\right\}$ of
expression~(\ref{br.2}). This optimization is implemented through the
adequate control over the car acceleration $a$, so, exactly the
acceleration plays the role of the control parameter available for the
driver actions.

A real driver can only approximately evaluate the quality of motion.
Let us describe the threshold in the driver perception of the motion
quality by
\begin{equation}
\mathcal{L}_{\text{thr}}=\frac{\lambda}{V}\epsilon_{c}^{2}
    \label{br.4}
\end{equation}
where $\epsilon_{c}$ is a small constant, $\epsilon_{c}\ll1$, because
definitely a driver can recognize the difference in the state of
staying and freely motion on an empty road.

When the controllable part $\mathcal{L}_{c}\left\{\mathfrak{h}\right\}
$ of the cost motion is much smaller then the threshold~(\ref{br.4})
the driver actually has no information of how to govern the car
motion. In this case it is natural to assume that he will not do
anything with respect to the car driving and, so, will fix the car
motion at the current state, including the current acceleration $a$.
Thereby the inequality
\begin{equation}
    \mathcal{L}_{c}\left\{\mathfrak{h}\right\}\lesssim\mathcal{L}_{\text{thr}}
    \label{br.5}
\end{equation}
determine the region in the phase space $\{h,v,a\}$ inside which the
driver cannot control the car motion. In evaluating such a driver
behavior with expression~(\ref{br.2}) we can formally treat $\delta$,
$\mathfrak{u}$, and $\mathfrak{a}$ as constants independent of one
another. So setting $\delta =h-h_{v}$, $\mathfrak{u}=v$, and
$\mathfrak{a}=a$ and taking into account the relation between $h_{V}$
and $V$ we get from condition~(\ref{br.5}) the approximate boundary of
this region in the following form
\begin{equation}
    \tau^{2}a^{2}+(v-V)^{2}+
    \frac{\Omega(h_V)}{4\tau^2}\frac{h_V}{h}(h-h_{V})^{2}
    \lesssim\epsilon_{c}^{2}\vartheta_{\text{max}}^{2}\,.
    \label{br.6}
\end{equation}
We recall that the parameter $g_h$ entering Eq.~(\ref{lge}) coincides
with $\sqrt{\Omega}/(2\tau)$. So we reproduced here the expression for
the rational driving boundary as has been introduced in
paper~\cite{BRDM} by a different line of reasonings.

\acknowledgments

These investigations were supported in part by RFBR
Grants~01-01-00389, 00439, UR Grant~01.03.005/2, INTAS Grant~00-0847,
and Russian Program ``Integration'' Project~B0056, during a stay of
one of the authors (IL) at the Institute for Transport Research of the
German Aerospace Centre.

%
\appendix
\section{Admissible family of the cost functions\label{Appp:A}}

A set $\mathbb{S}$ of one-dimensional paths $\left.\{x(t)\}\right\vert
_{t=0}^{\infty }$ of car motion is considered. All the paths are supposed to
originate from one point $x_0$ at the initial time $t=0$, i.e. $x(0)=x_0$, and
to go to infinity, $x(t)\rightarrow \infty $, as $t\to\infty$ such that the car
velocity $v(t)=dx/dt$ is bounded during the motion, $\left\vert v(t)\right\vert
<D$ (here $D$ is some constant).

It is assumed that the set $\mathbb{S}$ is ordered by a certain preference
relation ${}\preceq{}$ and there is a functional $\mathcal{L}\{x(t)\}$ of the
form
\begin{equation}
\mathcal{L}\{x(t)\}=\int_{0}^{\infty }dt\,w(x,t)\mathcal{F}(x,v,t)\,,
\label{appp1}
\end{equation}
measuring numerically the preference relation, i.e. meeting the condition
\begin{equation}
x_{1}(t)\preceq x_{2}(t)\quad \Leftrightarrow \quad \mathcal{L}
\{x_{1}(t)\}\geq \mathcal{L}\{x_{2}(t)\}\,.  \label{appp2}
\end{equation}
Here, first, the integrand $\mathcal{F}(x,v,t)$ depends on the car position on
the road $x(t)$, the velocity $v(t)$, and, may be, the current moment $t$ of
time. It plays the role of cost function, i.e. evaluates the quality of the car
motion state at current time. Second, the weight factor takes one of the forms
\[
w(x,t)=w_{a}(x)>0\quad \text{or}\quad w(x,t)=w_{b}(t)>0\,,
\]
where $w_{a}(x)\rightarrow 0$ and $w_{b}(t)\rightarrow 0$ as $x\rightarrow
\infty $ or $t\rightarrow \infty $. The functions $w_{a}(x)$ and $w_{b}(t)$
describe the spatial or temporal limitations in the driver prediction,
respectively. The examples
\[
w_{a}(x)=\exp \left( -\frac{x}{\lambda }\right) \quad \text{and}\quad
w_{b}(t)=\exp \left( -\frac{Vt}{\lambda }\right) \,,
\]
where $\lambda $ and $V$ are some constants, enable us to relate the cost
functional~(\ref{appp1}) to ones used in the main text. The specific form of
$w_{a}(x)$ and $w_{b}(t)$, however, does not affect the results to be obtained
in this Appendix.

Let $\widetilde{\mathcal{L}}\{x(t)\}$ be another cost functional of
form~(\ref{appp1}) with its own integrand $\widetilde{\mathcal{F}}(x,v,t)$ that
also describes the same preference relation.  The purpose of the present
Appendix is to derive the relationship between the cost functions
$\mathcal{F}(x,v,t)$ and $\widetilde{\mathcal{F}}(x,v,t)$.

For this purpose the path set $\mathbb{S}$ is divided into classes $\left\{
\mathfrak{S}\right\} $ of equivalent paths, i.e. $x_{1}(t),x_{2}(t)\in
\mathfrak{S}$ if $x_{1}(t)\sim x_{2}(t)$ or, what is the same, $\mathcal{L}
\{x_{1}(t)\}=\mathcal{L}\{x_{2}(t)\}$. The cost functional
$\widetilde{\mathcal{L }}\{x(t)\}$ based on the integrand
$\widetilde{\mathcal{F}}(x,v,t)$ has to preserve this class partition. So the
condition $\delta \widetilde{\mathcal{L}}/\delta x=0$ must hold for any
infinitesimal perturbation $\delta x(t)$ of a path $x(t)$ within the class
$\mathfrak{S}[x(t)]$.  Using the Lagrange multiplier technique and taking into
account the initial condition $x(0)=x_{0}$ (thus, $ \delta x(0)=0$) the latter
requirement is reduced to the equation
\begin{multline}
    \left( \Delta -v\partial _{v}\Delta \right) \partial _{x}w-
    (\partial _{v}\Delta)\partial _{t}w\\
    +  \left( \partial _{x}\Delta -\frac{d}{dt} \,\partial _{v}\Delta
    \right) w\,=0\,,
    \label{appp4}
\end{multline}
where the function $\Delta (x,v,t):=\widetilde{\mathcal{F}}(x,v,t)-\varkappa
\mathcal{F}(x,v,t)$ and $ \varkappa $ is a certain constant for all the points
$(x,v=dx/dt)$ on the phase plane $\left\{ x,v\right\} $ related by the given
curve $x(t)$. Let us show that the value of $\varkappa $ is constant for
\textit{all} the points of the phase plane. The cost functions
$\widetilde{\mathcal{F}}(x,v,t)$, $\mathcal{F}(x,v,t)$ are chosen, thereby, if
the \textit{quasi-local} equation~(\ref{appp4}) holds, the value of $\varkappa
$ must be determined by any fragment of the path $x(t)$ under consideration,
for example, by a fragment $x^{\ast }(t):=\left. x(t)\right\vert
_{t_{1}}^{t_{2}}\in \left. x(t)\right\vert _{0}^{\infty }$. For $t<t_{1}$ and
$t>t_{2}$ any point $(x,v)$ can be joined to the fragment $ x^{\ast }(t)$ by a
certain path $\bar{x}(t)$ which must be characterized by the same value of
$\varkappa $. The latter proves this statement because the time moments $t_{1}<
t_{2}$ have been taken arbitrary.

The cost function $\mathcal{F}(x,v,t)$ (as well as
$\widetilde{\mathcal{F}}(x,v,t)$) and the weight factor $w(x,t)$ describe
different effects of the car driving, so they are not related to each other.
Thus the terms of Eq.~(\ref{appp4}) containing $w$, $\partial _{t}w$, or
$\partial_{x}w$ should be equal to zero individually. Thereby
\begin{equation}
    \partial _{x}\Delta -\frac{d}{dt}\,\partial _{v}\Delta  =0
    \label{appp5}
\end{equation}
and
\begin{subequations}
\begin{align}
    \Delta -v\partial _{v}\Delta  & = 0\quad \text{for}\quad w(x,t)=w_{a}(x)\,,
    \label{appp6a} \\
    \partial _{v}\Delta  & = 0\quad \text{for}\quad w(x,t)=w_{b}(t)\,.
    \label{appp6b}
\end{align}
\end{subequations}
The solutions of system (\ref{appp5}), (\ref{appp6a}) and system (\ref{appp5}),
(\ref{appp6b}) have the forms
\[
\Delta (x,v,t)=C_{a}(x)\,v\quad \text{and}\quad \Delta (x,v,t)=C_{b}(t)\,,
\]
respectively, where $C_{a}(x)$ and $C_{b}(t)$ are arbitrary functions of $x$
and $t$. Thereby the desired relationship of the cost functions
$\mathcal{F}(x,v,t)$, $\widetilde{\mathcal{F}}(x,v,t)$ is given for $w(x,t)
=w_{a}(x)$ by the expression
\begin{subequations}
\begin{align}
    \widetilde{\mathcal{F}}(x,v,t) & = \varkappa \mathcal{F}(x,v,t)+
    C_{a}(x)\,v  \label{appp7a} \\
\intertext{and for $w(x,t) =w_{b}(t)$ by the expression}
    \widetilde{\mathcal{F}}(x,v,t) &= \varkappa \mathcal{F}(x,v,t)+
    C_{b}(t)\,.
    \label{appp7b}
\end{align}
\end{subequations}
where the constant $\varkappa$ must be positive, $\varkappa>0$.

Formulae (\ref{appp7a}) and (\ref{appp7b}) specify the admissible family of
cost functions entering cost functional~(\ref{appp1}) that preserve the
preference relation. As seen directly, integral~(\ref{appp1}) with the
integrands $C_{a}(x)$ and $C_{b}(t)$ is independent of a particular form of the
path $x(t)$. So the transformations~(\ref{appp7a}) and (\ref{appp7b})
correspond to the linear transformation of the cost functional.

\section{Lagrangian representation of the cost functional\label{app:B}}

The problem of finding the extremals of the functional
\begin{equation}
    L\{h(t)\}=\int_{0}^{\infty }dt\,w(t)\mathcal{F}(h,v,a)
    \label{appb.1}
\end{equation}
defined on a set of paths $\left. \{h(t)\}\right\vert _{t=0}^{\infty
}$ is considered. Here, first, the integrand $\mathcal{F}(h,v,a)$ is a
given function of the headway distance $h$, the current velocity
$v=V-dh/dt$, and the acceleration $a=-d^{2}h/dt^{2}$. Second, the
weight factor $\,w(t)>0$ is a function of time $t$ confined within
some time interval $(0,\tau _{V})$, i.e., $w(0)\sim 1$, $w(t)\ll 1$
for $t\gg \tau _{V}$, and $w(t)\rightarrow 0$ sufficiently fast as
$t\rightarrow \infty $. So, $w(0)=1$ can be adopted without lost of
generality. Third, the trial paths $\{h(t)\}$ meet the initial
conditions
\begin{equation}
    h(0)=h_{0}\,,\quad v(0)=v_{0}\,,
    \label{appb.2}
\end{equation}
and do not diverge as time goes on, i.e., there is a number $C$ so
that
\begin{equation}
    \left\vert h(t)\right\vert <C\,.
    \label{appb.3}
\end{equation}

The properties of the functional extremals have been studied already.
Keeping in mind the results in Sec$.~$\ref{sec:2}, a special case is
analyzed in the following. Let $\{\tau _{d}\}$ be the characteristic
time scale of the velocity relaxation to the stationary value $V$. In
particular, the values of $\{\tau _{d}\}$ are estimated by the
eigenvalues of the corresponding extremal equation linearized near the
stationary point. The case $\tau _{d}\ll \tau _{V}$ is considered here
to show how the functional~(\ref{appb.1}) can be rewritten to
eliminate the time dependent factor $w(t)$. In this way the problem of
finding the extremals gets the classical form met in theoretical
mechanics.

The stationary point $\{h_{V},V,a=0\}$ is determined by the properties
of both the function $\mathcal{F}(h,v,a)$ and the factor $w(t)$ (see
Exp.~(\ref{2.11})). As a result, at the stationary point (see again
Exp.~(\ref{2.11}))
\begin{equation}
    \partial _{a}\mathcal{F}_{\text{st}}=0\,,\quad \partial _{h}\mathcal{F}_{
    \text{st}}\propto \frac{1}{\tau _{V}}\,.
    \label{appb.4}
\end{equation}
The derivative $\partial _{v}\mathcal{F}_{\text{st}}$ does not contain
factors similar to $\tau _{d}/\tau _{V}$. So, $w(t)$ cannot be omitted
directly. As a first step the function $\mathcal{F}(h,v,a)$ is
replaced by the difference $\Delta \mathcal{
  F}(h,v,a):=\mathcal{F}(h,v,a)-\mathcal{F}(h_{V},V,0)$. This does not
affect the extremals. The difference $\Delta \mathcal{F} (h,v,a)$ as a
function of $t$ is practically confined within a time interval
$(0,T_{d})$ whose upper boundary $T_{d}\gtrsim \max \{\tau _{d}\}$
and, so, $T_{d}\ll \tau _{V}$.  Therefore, at the next step
expression~(\ref{appb.1}) is rewritten as
\begin{multline}
    L\{h(t)\}\cong \int_{0}^{\infty }dt\,\Big\{\Delta \mathcal{F}(h,v,a)
    \\
    -\frac{t}{\tau _{V}}(v-V)\partial _{v}\mathcal{F}(h_{V},V,0)\Big\}
    \label{appb.5}
\end{multline}
which is justified to the first order in the small parameter $\max
\{\tau _{d}\}/\tau _{V}$ by virtue of expressions~(\ref{appb.4}) and
the fact that the second term in Exp.~(\ref{appb.5}) plays a
substantial role only when the path $h(t)$ tends to the stationary
point. Here the symbol $\cong$ means that functional~(\ref{appb.5})
possesses the same collection of the extremals as the initial
functional $L\{h(t)\}$ within the adopted accuracy and the time
dependent factor $w(t)$ has been approximated by
\begin{equation}
    w(t)\approx 1-\frac{t}{\tau _{V}}\,.
    \label{appb.6}
\end{equation}
Formula~(\ref{appb.6}) can be regarded actually as the definition of
the time scale $\tau_{V}:=\left[ -dw(0)/dt\right] ^{-1}$. Taking into
account the relation $ (v-V)=-dh/dt$ and integrating the second term
in Exp.~(\ref{appb.5}) by parts yields
\begin{multline}
    L\{h(t)\}\cong \int_{0}^{\infty }dt\,\Big\{ \Delta \mathcal{F}(h,v,a)\\
    -\frac{(h-h_{V})}{\tau _{V}}\partial _{v}\mathcal{F}(h_{V},V,0)\Big\} \,.
    \label{appb.7}
\end{multline}
Omitting the constant components of the integrand which does not
affect the extremals the required result follows
\begin{equation}
L\{h(t)\}\cong \int_{0}^{\infty }dt\,\Big\{ \mathcal{F}(h,v,a)-\frac{h}{\tau
_{V}}\partial _{v}\mathcal{F}(h_{V},V,0)\Big\} \,.  \label{appb.8}
\end{equation}
Minimization of functional~(\ref{appb.8}) gives the desired extremals
and integral~(\ref{appb.8}) does not contain a time dependent factor.
The function
\begin{equation}
    \mathcal{F}_{\text{eff}}(h,v,a\mid V):=\mathcal{F}(h,v,a)-\frac{h}{\tau _{V}}
    \partial _{v}\mathcal{F}(h_{V},V,0)
    \label{appb.9}
\end{equation}
can be regarded as a Lagrangian of the car dynamics.
%

\end{document}